\documentclass{article}

\PassOptionsToPackage{numbers, compress}{natbib}
\usepackage[final]{neurips_2019}

\usepackage[utf8]{inputenc} 
\usepackage[T1]{fontenc}    
\usepackage{hyperref}       
\usepackage{url}            
\usepackage{booktabs}       
\usepackage{amsfonts}       
\usepackage{nicefrac}       
\usepackage{microtype}      

\usepackage{graphicx}
\usepackage{dcolumn}
\usepackage{bm}
\usepackage{bbold}
\usepackage{braket}
\usepackage{mathtools}
\usepackage{setspace}

\title{Nanophotonic Inverse Design with SPINS: Software Architecture and Practical Considerations}

\author{
Logan Su$^1$,
Dries Vercruysse$^1$,
Jinhie Skarda$^1$,
Neil V. Sapra$^1$,\\
\textbf{
Jan A. Petykiewicz$^1$,
Jelena Vu\v{c}kovi\'c$^1$}\\
\\[0.5em]
$^1$E. L. Ginzton Laboratory, Stanford University, Stanford, CA 94305, USA \\
\texttt{logansu@stanford.edu}
}

\begin{document}

\maketitle

\begin{abstract}
A computational nanophotonic design library for gradient-based optimization called SPINS is presented. Borrowing the concept of computational graphs, SPINS is a design framework that emphasizes flexibility and reproducible results. The mathematical and architectural details to achieve these goals are presented, and practical considerations and heuristics for using inverse design are discussed, including the choice of initial condition and the landscape of local minima.
\end{abstract}

\section{Introduction}
Photonics has many applications ranging from optical interconnects \cite{sun2015single} to augmented reality (AR) \cite{kress2013review} to optical neural networks \cite{shen2017deep}. The wide variety of applications necessitates a wide variety of building blocks that compose photonic systems. A large body of research is devoted to optimizing these photonic structures using simple geometries where the physics is well-understood. The devices are then further optimized via parameter sweeps or genetic or particle swarm optimization over small number of degrees of freedom. Though simple, these design approaches only explore a small fraction of the possible designs.

Computational nanophotonic design using gradient-based optimization, also known as inverse design, is a promising method that provides an efficient mechanism through which to explore the full space of possible designs. The number of degrees of freedom in a typical design space is so large that it is infeasible to simulate even a small fraction of the possible designs. This can be mitigated by using the gradient (i.e. sensitivity of the loss function with respect to changes in the permittivity distribution) to guide the optimization process. Because the gradient can be computed with only one additional electromagnetic simulation, this {\it adjoint method} is computationally efficient and has been used to demonstrate devices that have smaller footprints, better efficiencies, and novel functionalities \cite{molesky2018inverse}.

In this paper, we present SPINS (Stanford Photonic INverse design Software) \cite{spins-b, spins}, a computational nanophotonic design framework for running gradient-based optimization that has been used to optimize devices in several previous works \cite{dory2019inverse,vercruysse2019analytical,yang2019inverse,sapra2019chip,sapra2019inverse,su2018fully}. SPINS is a design framework, not a design methodology: It is a way to formulate and express the optimization problem. That is, SPINS is agnostic to specific optimization procedure or type of nanophotonic problem (e.g. waveguide-based, free space photonic elements). Instead, SPINS uses software building blocks called {\it nodes} that are assembled together into a {\it problem graph}. This enables easy experimentation during the design process as well as a way to customize and extend the functionality of SPINS. The problem graph can also be saved and restored, a feature that is invaluable, for example, in reproducing results or in restarting optimizations.

Using inverse design effectively also requires an understanding of the ``control knobs" available to the designer. To that end, we discuss practical considerations and heuristics when using inverse design. These include the design of objective functions and the choice of initial condition based on an analysis of the local minima reached by the optimization process. 

The paper outline is as follows. Section \ref{sec:overview} provides a mathematical overview of inverse design, and Section \ref{sec:framework} provides an overview of the framework and how the inverse design formulation is implemented. Section \ref{sec:wdm-example} presents an example of designing wavelength demultiplexers and discusses salient points in the overall design process. Finally, Section \ref{sec:analysis} discusses practical considerations and analyzes key properties of gradient-based nanophotonic optimization.

\section{Inverse Design Formulation Overview}
\label{sec:overview}
This section describes the mathematical foundations behind inverse design approach used in SPINS. Although the exact form of the optimization problem varies from device to device, photonic design problems generally share a common set of features, which are described below. Note that in this manuscript, all variables represent finite-dimensional vectors and therefore represent discretized quantities. More details can be found in Appendix \ref{sec:app-math-details}.


\subsection{General Optimization Problem}
\begin{figure}
    \centering
    \includegraphics[width=2.8in]{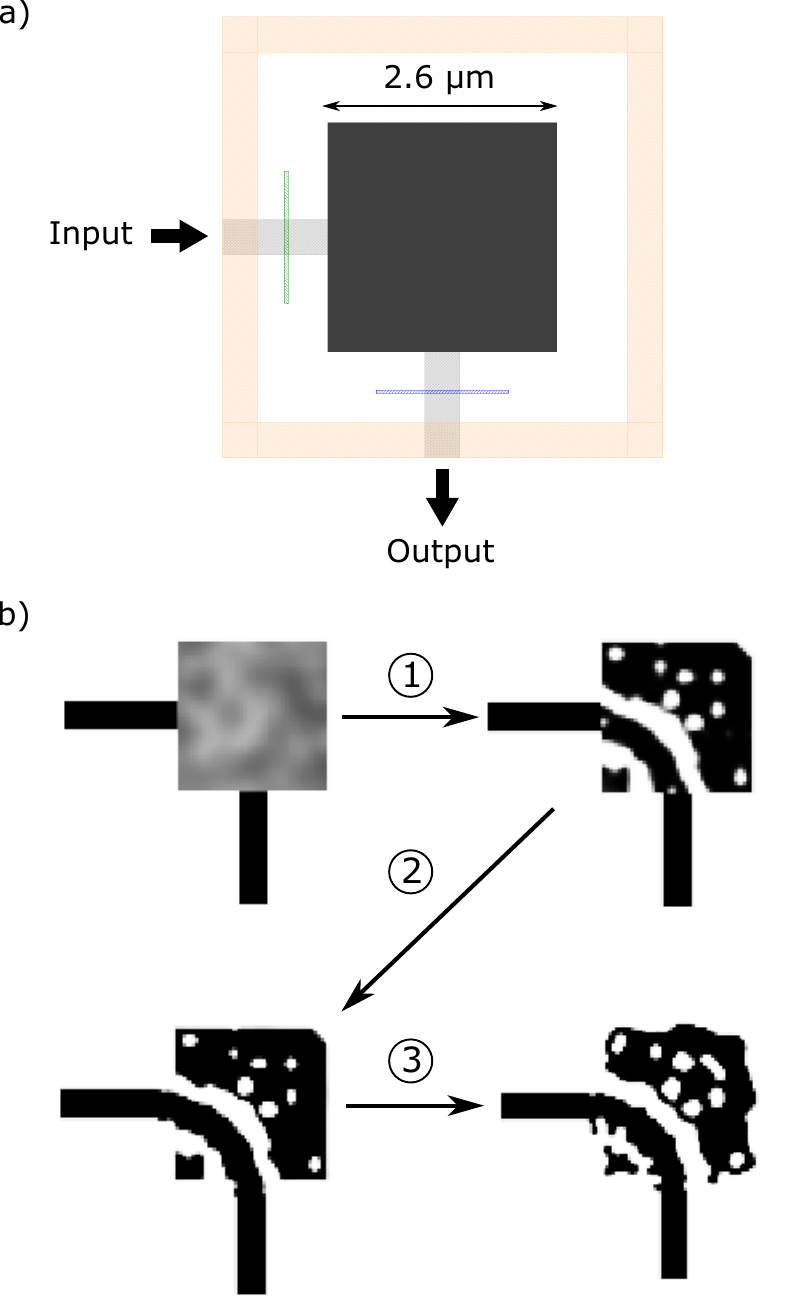}
    \caption{a) Problem setup for a waveguide bend. The orange frame represents the simulation domain, the gray boxes denote the input and output waveguides, and the dark gray square represents the design area, the region in which the permittivity distribution is allowed to vary. The input mode is injected at the location designated by the green rectangle, and the modal overlap with the fundamental mode is computed at the blue rectangle. b) Optimization sequence. First, randomly-generated initial structure with continuous permittivity distribution is optimized (continuous optimization). Second, the resulting continuous structure is converted into a discrete one (discretization). Third, the discrete structure is further optimized (discrete optimization). Fabrication constraints are enforced at this time.}
    \label{fig:wgbend}
\end{figure}
Many photonic design problems can be cast into the following form:
\begin{equation}
	\label{eqn:opt-prob}
	\begin{split}
	\min_{p} \quad & f_{obj}(E(\epsilon(p))) \\
	\textrm{subject to} \quad &  p \in S_{fab}
	\end{split}
\end{equation}
where $f_{obj}$ is the objective function that captures the goal of the optimization,  $E$ is the electric field distribution, $\epsilon$ is the permittivity distribution that is described by a parametrization vector $p$, and $S_{fab}$ is the set of fabricable devices. For example, for the optimization of a waveguide bend (Figure \ref{fig:wgbend}a), $f_{obj}$ is the negative of the transmission efficiency (since the objective is to be minimized), and $S_{fab}$ corresponds to all devices with minimum feature size greater than 100 nm. Not all photonic design problems can be cast into the exact form given by Problem \ref{eqn:opt-prob}, but they generally take on a similar form. For instance, to optimize for temperature insensitivity, a possible problem is:
\begin{equation}
\begin{split}
	\min_{p} \quad & f_{obj}(E(\epsilon_1(p))) + f_{obj}(E(\epsilon_2(p)))\\
	\textrm{subject to} \quad &  p \in S_{fab}
	\end{split}
\end{equation}
where $\epsilon_1$ and $\epsilon_2$ correspond to the permittivity distribution at different temperatures.

\subsection{Parametrization $\epsilon(p)$}
\label{sec:overview-parametrization}
In principle, it is possible to optimize the permittivity distribution directly (i.e. control the value of permittivity distribution in every point in space). However, because of fabrication constraints, it is often more convenient to indirectly specify the permittivity distribution via a {\it parametrization}.

A fundamental difficulty in photonic design is that arbitrary permittivity distributions cannot be fabricated. First, devices are usually composed of a small number of distinct materials, so the permittivity can only take on certain discrete values. Second, devices are often fabricated with top-down lithography, so the permittivity along the vertical direction must be the same. Last, devices usually have minimum size feature constraints.

By choosing an appropriate parametrization, these fabrication constraints can be more naturally imposed. For example, the 1D grating coupler design parametrization used in \cite{su2018fully} defines the elements of $p$ as the distance between grating edges. This way, fabrication constraints are specified by constraining the minimum distance between neighboring grating edges. For 2D designs, levelset parametrizations are commonly used in SPINS, which naturally define binary devices (Section \ref{sec:levelsets}).

\subsection{Simulation $E(\epsilon)$}
The electromagnetic simulation can be considered a function that accepts a permittivity distribution and outputs the electric fields. For example, in the frequency domain, the electric field is computed from the permittivity by inverting Maxwell's equations:
\begin{align}
    E = \left(\left(\nabla \times \frac{1}{\mu} \nabla \times\right) - \omega^2 \epsilon(p)\right)^{-1} (-i\omega J)
\end{align}
where $J$ is the input source and $\omega$ is the frequency. In Figure \ref{fig:wgbend}a, $J$ injects the fundamental mode into the input waveguide. Note that a time-domain problem (e.g. pulse-shaping) can be formulated in terms of a frequency-domain problem by taking the Fourier transform. It is also possible to extend the formulation to nonlinear devices \cite{hughes2018adjoint} and eigenmode problems \cite{wang2011robust}. 

\subsection{Objective Function $f(E)$}

The form of the objective function varies depending on the optimization goal, and consequently, it is one of the important control knobs for a designer. For example, a possible objective for maximizing transmission is:
\begin{align}
    f_{obj}(p) = -|c^\dagger E(p)|^2
\end{align}
where $c^\dagger E$ computes the modal overlap of the electric field $E$ with the target mode $c$ at the output port of the device. In Figure \ref{fig:wgbend}, this would correspond to measuring the modal overlap with the fundamental mode at the bottom waveguide. If the goal is to have transmission at a particular value, a possible objective is
\begin{align}
     f_{obj}(p) = (t -|c^\dagger E(p)|)^2
\end{align}
where $t$ is the target transmission. This can be extended to handle multiple sub-objectives. For example, an objective for multiple frequency optimization (e.g. for wavelength demultiplexing or broadband problems) could be:
\begin{align}
f_{obj}(p) =  f_{obj, 1}(E_1(\epsilon_1(p))) + f_{obj, 2}(E_2(\epsilon_2(p)))
\end{align} In such cases, the form of the objective function plays a critical role in determining the trade-off between the sub-objectives. Discussion of this trade-off and other common objective functions are described in Section \ref{sec:objective-funs}.

\subsection{Solving the Design Problem}

\label{sec:overview-transformations}
Generally, directly optimizing for a discrete device with fabrication constraints produces poor devices because the landscape is highly non-convex and a good initial condition is required. Initializing with a classically-designed device is possible, but this approach limits the design space. Alternatively, rather than solving the optimization problem in a single pass, the optimization can be broken down into a series of sub-optimizations (Figure \ref{fig:wgbend}b). One approach is a continuous relaxation whereby the permittivity is allowed to vary continuously between the device and cladding. A discretization operation then converts the device at the end of this {\it continuous stage} into a discrete device for further optimization.

Possible discretization techniques include thresholding and minimization of the difference between continuous and discrete structures (also see Section \ref{sec:discretization}). In thresholding, a pixel in the discrete structure is considered to be 1 if the continuous structure greater than 0.5 and a pixel is 0 if the continuous structure is less than 0.5. Though simple to implement, this method usually does not provide the best results. Instead, discretization in SPINS often involves solving an optimization problem subject to any fabrication constraints:

\begin{equation}
    \underset{p}{\text{min }} || \epsilon_{disc}(p) - \epsilon_{cont} ||
\end{equation}

where $\epsilon_{cont}$ is the continuous permittivity and $\epsilon_{disc}$ is the discrete permittivity. Since this optimization problem does not depend on the field, it is quick to optimize, though the method with which to solve the problem is dependent on the choice of parametrization $p$ \cite{vercruysse2019analytical, su2018fully}.

In general, each of the optimization stages can be considered as a {\it transformation} that converts one parametrization into another. Some transformations (e.g. discrete optimization) preserve the parametrization type but change the values whereas other transformations (e.g. discretization) actually change the parametrization type. In effect, solving the design problem involves choosing a sequence of transformations that starts with a randomized structure and produces a fully-optimized one.

\section{Framework Overview}
\label{sec:framework}
Though the exact formulation of a design problem varies widely from device to device, most design problems share a similar structure, requiring some form of parametrization, simulations, and mathematical functions to construct the objective. There are many fundamental building blocks that are shared amongst different design problems, such as computation of an overlap of the form $|c^\dagger E|$. Consequently, SPINS is structured around small building blocks that can be assembled together to formulate any desired design problem.

\begin{figure}
    \centering
    \includegraphics[width=3in]{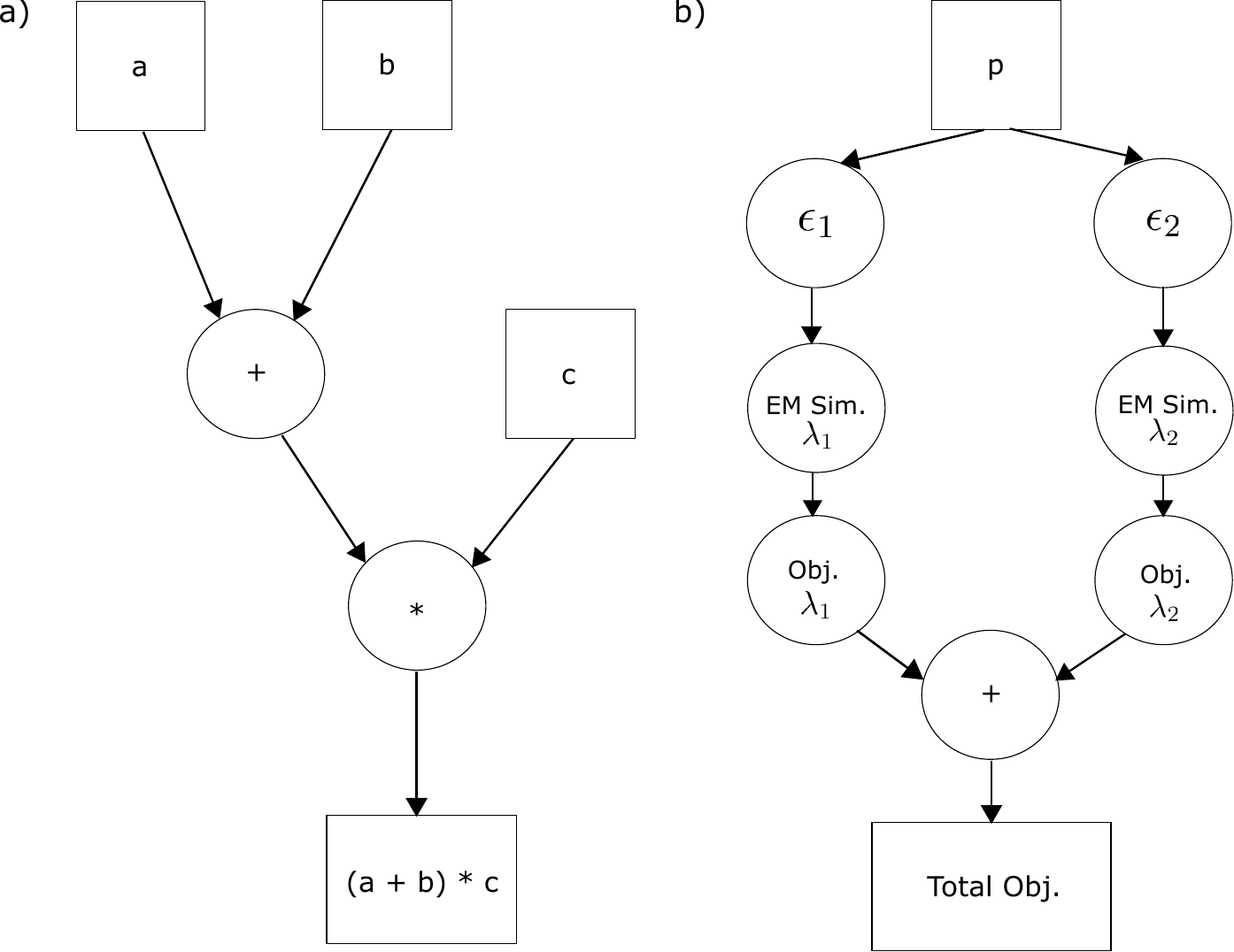}
    \caption{Illustrations of the computational graph. Boxed elements correspond to inputs and outputs of the graph, whereas circles correspond to operations. a) An example of computational graph that accepts as input three values $a$, $b$, and $c$ and computes as output $(a + b) \times c$. b) Schematic computational graph corresponding to EM objective with two different wavelengths. First, the parametrization vector is used to compute permittivity distribution at the two wavelengths (in order to incorporate any material dispersion). The EM simulation operation accepts the permittivity distributions and produces the electric field as output, which is then used to compute an objective function for the particular wavelength (e.g. power at an output port). The sub-objectives are aggregated together (in this case with a sum) to produce the final objective function value.}
    \label{fig:problem-graph}
\end{figure}

SPINS is built around the notion of an {\it optimization plan} consisting of two parts. First, fundamental building blocks called {\it nodes} are assembled together to form a {\it problem graph}. The problem graph provides a complete description of the design problem, from the details of the simulation to the precise form of the objective function (Figure \ref{fig:problem-graph}). Second, there is a sequence of {\it transformations} that define the optimization strategy. Transformations either modify the values of parametrization or convert one form of parametrization into the other. These transformations include continuous and discrete optimization stages as well as the discretization process. The transformations use the problem graph to compute any necessary quantities, such as the objective function value.

The problem graph is merely a description of the problem and does not contain actual implementation. Consequently, no computation is performed when setting up the graph. This is similar to static computational graphs in TensorFlow \cite{tensorflow2015-whitepaper} and carries many important advantages in inverse design:

\begin{itemize}
    \item The user can easily experiment with different types of objective functions by assembling simple functions together like building blocks. Moreover, design blocks created by other users can be easily shared with others.
    \item The gradients can be automatically computed via reverse-mode autodifferentiation \cite{rumelhart1988learning}, obviating the need to manually implement gradient calculations.
	\item It provides a record of the exact optimization sequence as well as the hyperparameters used to run the optimization. Hyperparameters are often critical in producing good results, and this provides a mechanism by which to store the exact values used to produce a particular result. Importantly, during post-optimization analysis, the values of any node in the graph can be determined, even if not explicitly saved.
	\item Optimizations can be resumed from any iteration. This is useful for resuming long optimizations that were stopped (e.g. due to hardware failure), or for experimenting with different optimization sequences (e.g. swapping out the discrete optimization or adding another discrete transformation).
	\item Optimization problems can be generated in one location and executed on a remote server, as the whole problem is specified through a JSON file and GDS files. Therefore, in principle, the optimization problem generation does not even require programming knowledge, enabling designers to work independently from software developers.
	\item Similarly, optimization problems can be generated at one time and run later, which is useful for queueing up optimization problems to run in batches for sweeps.
\end{itemize}

The typical workflow with SPINS involves 1) setting up the problem graph, 2) choosing the transformations, and 3) executing the optimization plan. SPINS is designed to allow users to add custom nodes and transformations to suit their needs. Depending on the outcome of the optimization, the designer may opt to experiment with different objective functions or sequence of transformations, which is enabled by the mix-and-match capability of SPINS parametrizations, nodes, and transformations.

The outline for a typical problem graph for a nanophotonic design problem is depicted in Figure \ref{fig:problem-graph}b. The parametrization is the primary input to the graph. The permittivity function converts the parametrization into an actual permittivity distribution, which is then fed to the simulation function. The simulation function outputs the corresponding electric fields, which is used to feed a subgraph that produces the objective function value. If there are multiple sub-objectives, these sub-objectives are combined together to form one final total objective function. Appendix \ref{sec:app-frame-details} discusses the implementation of each of these components.

\section{Wavelength Demultiplexer Design Examples}
\label{sec:wdm-example}
In this section, we use SPINS to design wavelength demultiplexers in 3D. This is a canonical benchmarking problem for inverse design because classical designs at these small footprints do not exist, and randomly choosing a structure results in a non-functioning device. To illustrate practical concerns that arise during optimization as well as the importance of flexibility in a design framework, three different demultiplexers are designed: a continuous structure to estimate the maximum possible performance; a discrete structure that can be fabricated; and a structure that controls both the amplitude and phase at the output ports. 

\begin{figure*}
    \centering
    \includegraphics[width=\textwidth]{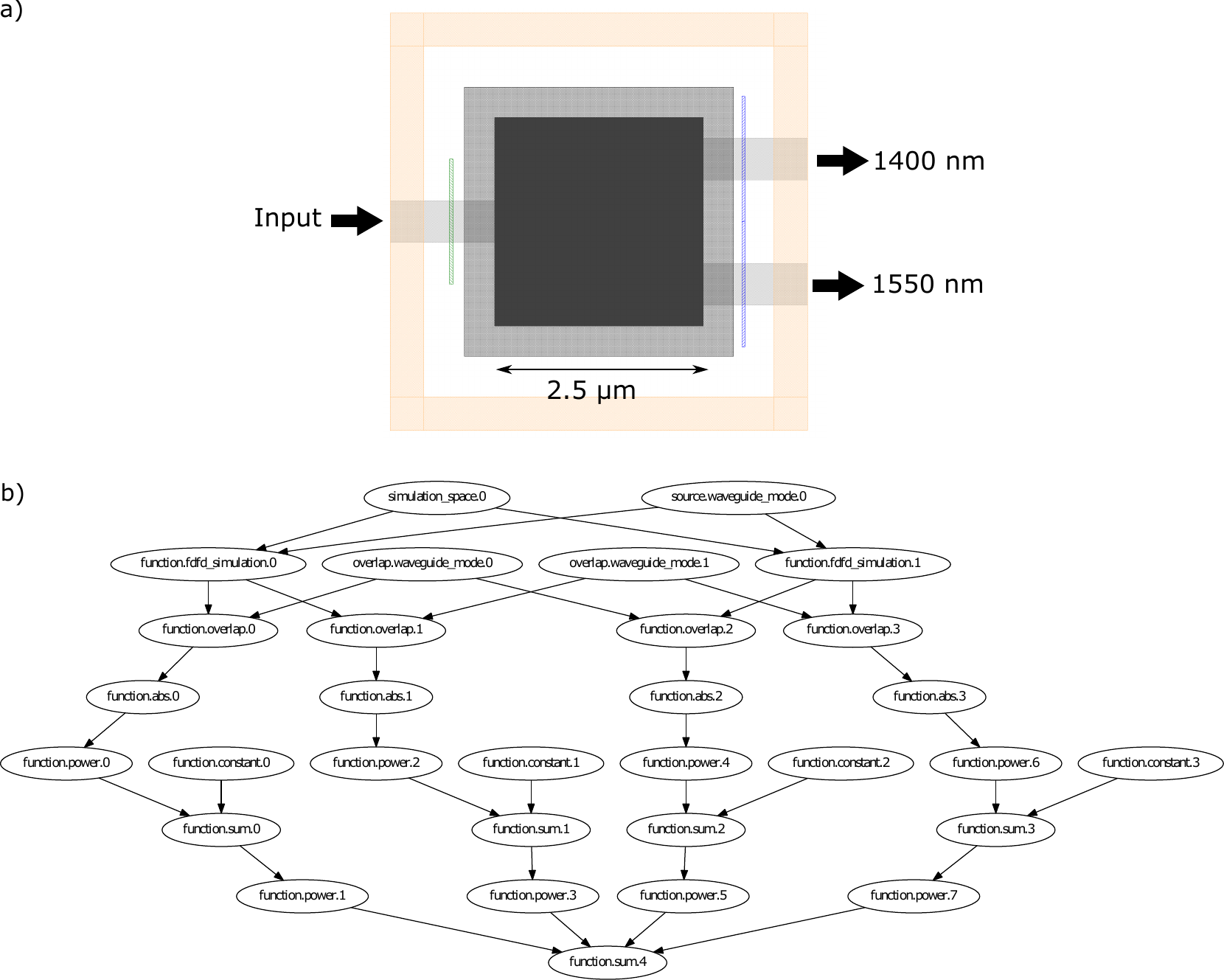}
    \caption{a) Problem setup for the wavelength demultiplexer. The goal is route to 1400 nm through the upper waveguide and 1550 nm through the bottom waveguide. b) Problem graph for objective defined in Section \ref{sec:wdm-example-obj}.}
    \label{fig:wdm-example}
\end{figure*}

\subsection{Simulation Space Setup}
We design a 3D device for the 220 nm SOI platform ($n = 3.5$ for illustration purposes) with a 2.5 um x 2.5 um design area and 400 nm input and output waveguides. The goal is to route 1400 nm to the upper waveguide and 1550 nm to the lower waveguide.

Figure \ref{fig:wdm-example}a shows the simulation setup from a top-down view of the device. The dark gray square represents the design area for the continuous stage transformations in which the permittivity is allowed to change. When optimizing a discrete device, the design region indicated by the the light gray square surrounding the dark gray square is used. The discrete stage has an enlarged design area compared to the continuous transformation so that the discrete optimization can ensure a smooth transition between the waveguides outside the design area to the structure within the design area. Without the enlarged design region, it is possible for fabrication constraints to be violated at the boundary between the waveguides and the design area.

 The rest of the simulation setup is similar to the setup for a normal electromagnetic simulation. The light gray rectangles protruding from the design area indicate the input and output waveguides. These extend into the yellow region denoting the perfectly matched layer (PML) boundary condition used in the simulation. The green rectangle indicates the location of the mode source used to inject the fundamental mode into the input waveguide, and the blue rectangles denote the location of the modal overlaps used to compute the power injected into the output waveguides. The values of these modal overlaps are used to compute the overall objective function.

\subsection{Objective Function}
The optimization problem is defined as
\label{sec:wdm-example-obj}
\begin{equation}
\begin{split}
\min_{p,E_{1}, E_{2}} \quad & f(|c_{t, 1}^\dagger E_{1}|^2, 1) + f(|c_{t, 2}^\dagger E_{2}|^2, 0)  \\
& + f(|c_{b, 2}^\dagger E_{2}|^2, 1) + f(|c_{b, 1}^\dagger E_{1}|^2, 0) \\
\textrm{subject to} \quad & \nabla \times \frac{1}{\mu} \nabla \times E_{1} - \omega_1^2 \epsilon(p) E_{1} = -i\omega_{1} J_{1} \\
& \nabla \times \frac{1}{\mu} \nabla \times E_{2} - \omega_2^2 \epsilon(p) E_{2} = -i\omega_{2} J_{2} 
\end{split}
\end{equation}
where $\omega_n$ represent the frequencies at 1400 nm and 1550 nm; $E_n$ is the electric field for frequency $\omega_n$; $J_{n}$ injects a TFSF source into the input waveguide for frequency $\omega_n$; $c_{a,n}$ is the overlap vector for frequency $\omega_n$ for the top ($a = t$) or bottom ($a=b$) output waveguide such that $|c_{a,n}^\dagger E_{n}|^2$ is the power going into the fundamental mode of the output waveguide; and $f(x, y) = (x - y)^2$. The problem graph corresponding to this objective is shown in Figure \ref{fig:wdm-example}b.

The objective is a sum with four terms, with each term corresponding to a sub-objective. Two terms ($s = 1$) correspond to maximizing transmission through the top waveguide at 1400 nm and transmission through the bottom waveguide at 1550 nm, and two terms ($s = 0$) correspond to rejection modes, i.e. discourage 1550 nm at the top waveguide and 1400 nm at the bottom waveguide. Without the rejection terms, the crosstalk could be higher. These four sub-objectives can compete with one another. For instance, at any given optimization step, it could be possible to increase transmission at 1400 nm through the top waveguide but at the cost of higher crosstalk at 1550 nm (transmission of 1550 nm through the same waveguide). Consequently, it is important to balance out the objectives as desired through the construction of the objective function. In this particular case,
$f(x, y)$ squares the difference in order to prefer devices that are more balanced, i.e. a device with equal transmission at 1400 nm and 1550 nm is preferable to one where there is high transmission at 1400 nm but low transmission at 1550 nm.

One may also want to include other sub-objectives. For example, an additional term could be included to minimize back-reflection at the input waveguide. One common extension is to optimize for broadband wavelength dependence, which can be achieved in this case by simulating at additional wavelengths near 1400 nm and 1550 nm (see Section \ref{sec:objective-funs-broadband}). Moreover, broadband behavior can be used as a crude proxy for robustness as well. Since a small perturbation in either refractive index or structure results in a spectral shift, broadband optimization can make the structure more robust to temperature and fabrication errors.

\begin{figure*}
    \centering
    \includegraphics[width=4.5in]{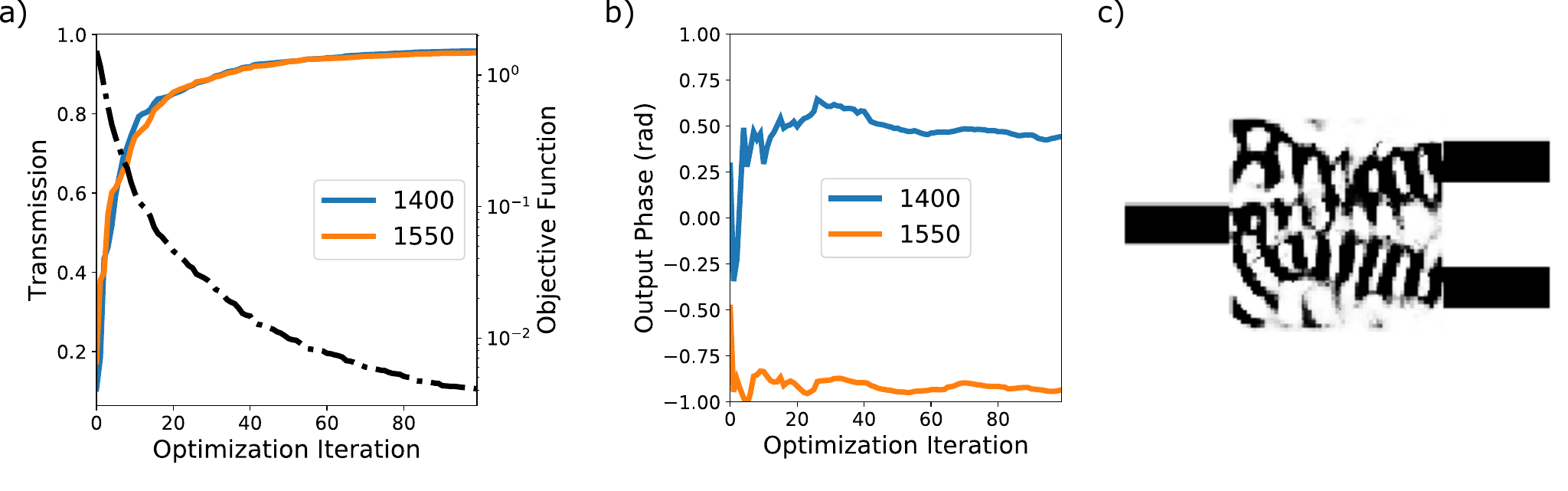}
    \caption{Device A1. a) Transmission (solid) and objective value (dotted) as a function of optimization iteration.  b) Phase of the output waveguide mode as a function of optimization iteration. c) Final device design.}
    \label{fig:wdm-device-1a}
\end{figure*}

\subsection{Device A1: Continuous Device for Estimating Performance}

\label{sec:wdm-continuous}
Before designing a fabricable device, it is helpful to check that the given design area is large enough to achieve desired performance targets. To do so, we use a {\it direct parametrization} where the permittivity at every pixel of the device area can be controlled independently and can take on a continuous value. Because the design space for this problem is much larger than the design space of fabricable devices, the outcome of this optimization provides an estimate of the upper bound on the performance of a device with this design area.

Figure \ref{fig:wdm-device-1a}c shows the final device structure for this optimization. The device has a transmission of 95\% at both 1400 nm and 1550 nm with over 19 dB crosstalk suppression. The device was optimized for 100 iterations and took roughly 3.5 hours running with 4 Titan Black GPUs and 2 CPUs with 7.5 GB RAM.

The optimization trajectory of the device is shown in Figure \ref{fig:wdm-device-1a}a. The objective function always decreases because the optimizer used employs a line search in which a step is taken only if the objective function decreases. However, the amount of improvement drops dramatically after tens of iterations. In fact, the optimization has mostly converged by iteration 30, with the remaining 70 iterations improving the transmission by around 5\%. Consequently, when used to estimate device performance, it often suffices to optimize for only 20-30 iterations. If the desired performance targets are too far from being met after a few tens of iterations, then a larger design area should be probably be considered.

\subsection{Device A2: Fabricable Discrete Device}

\begin{figure*}
    \centering
    \includegraphics[width=3.5in]{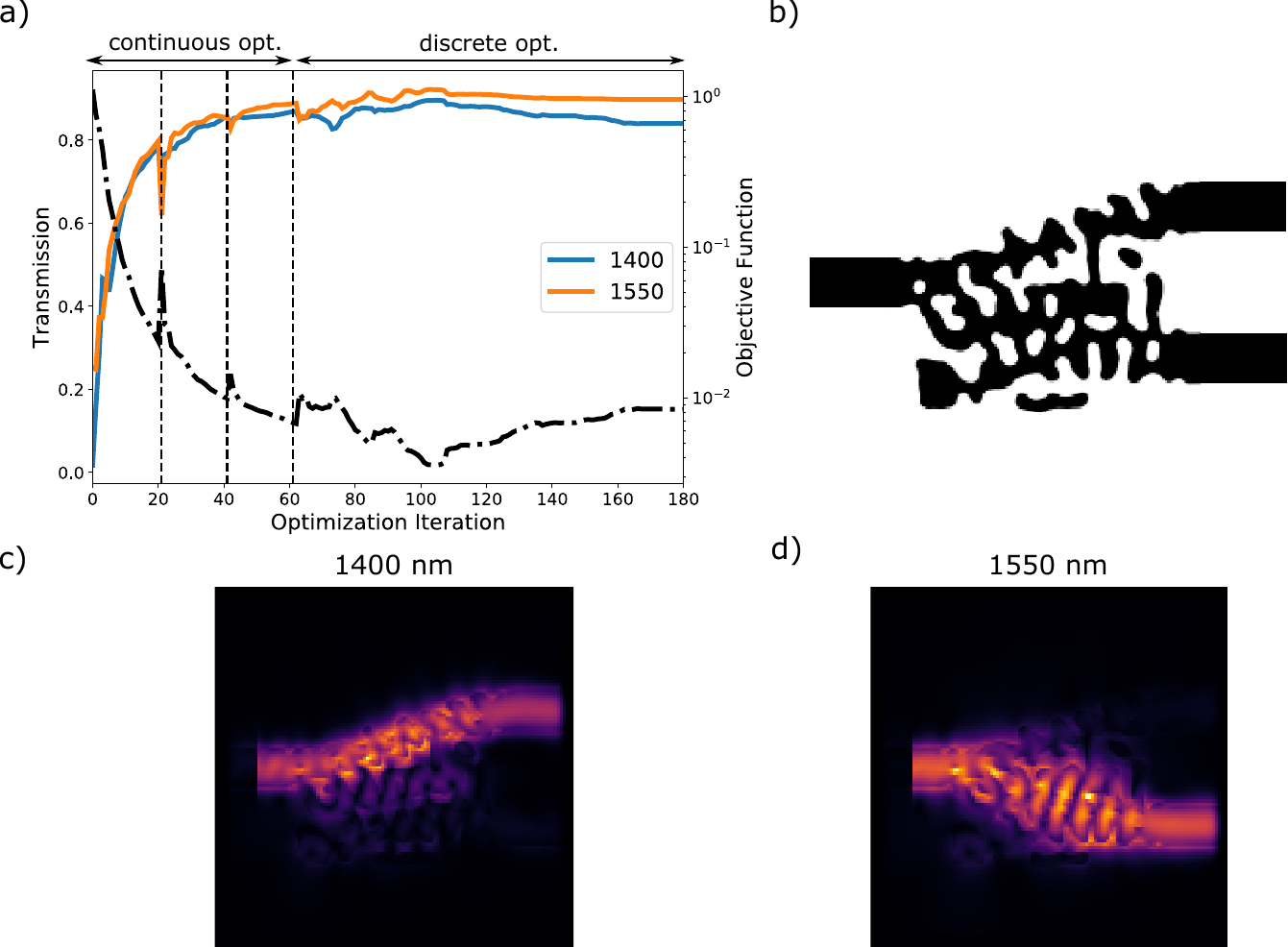}
    \caption{Device A2. a) Transmission (solid) and objective value (dotted) as a function of optimization iteration. The vertical dotted lines indicate the start of a new transformation. Discretization occurs in between the continuous and discrete optimization stages. b) Final device design. c) Electric field intensity at 1400 nm. d) Electric field intensity at 1550 nm.}
    \label{fig:wdm-device-1b}
\end{figure*}

\label{sec:wdm-discrete}
To optimize a fabrication-constrained device, the standard continuous relaxation approach discussed in Section \ref{sec:overview-transformations} was used. This involves adding transformations to handle discretization and the discrete optimization. In addition, a cubic interpolation parametrization is used instead of a direct parametrization in the continuous optimization stage in order to ensure better discretization for the continuous structure \cite{vercruysse2019analytical}. During the discrete optimization stage, a constraint was added to ensure devices with feature sizes greater than 100 nm and radius of curvature greater than 50 nm result from the discrete transformation.

The optimization trajectory is plotted in Figure \ref{fig:wdm-device-1b}a. The dotted lines indicate different optimization transformations. In this case, the continuous stage actually consists of three optimization transformations, whereas the discrete stage consists of a single transformation. In the continuous stage, each of the transformations bias the structure to become a little more discrete. As before, in the continuous transformations, the objective function always decreases except at the transition between transformations, at which point a parameter is adjusted to make the structure more discrete \cite{vercruysse2019analytical}. After the three continuous transformations, the discretization transformation changes the structure to be fully discrete, resulting in another increase in objective function. In the discrete transformation, the objective value decreases and increases because the fabrication constraint penalty is gradually increased throughout the transformation. Eventually, the optimization recovers the performance achieved in continuous stages, indicating that discretization and discrete optimizations worked sufficiently well.

Note that the continuous stage performs worse than the continuous stage achieved in Section \ref{sec:wdm-continuous}. The reasons are two-fold. First, the cubic interpolation parametrization factors in the desired feature size to avoid small features and is therefore more restrictive than the direct parametrization. Second, the continuous optimization was terminated before reaching convergence. As a result, at some iterations, even the discrete optimization achieved a better performance than the continuous optimization. 

Figure \ref{fig:wdm-device-1b}b shows the final device design and Figures \ref{fig:wdm-device-1b}c-d show the electric field distributions at 1400 nm and 1550 nm. The device has a transmission of 84\% dB at 1400 nm and 90\% at 1550 nm with crosstalk suppression of over 20 dB. Overall the device took roughly 11 hours long to optimize running with 2 K80 GPUs and 2 CPUs with 7.5 GB RAM.

\subsection{Device B: Phase Control}
\begin{figure*}
    \centering
    \includegraphics[width=4.5in]{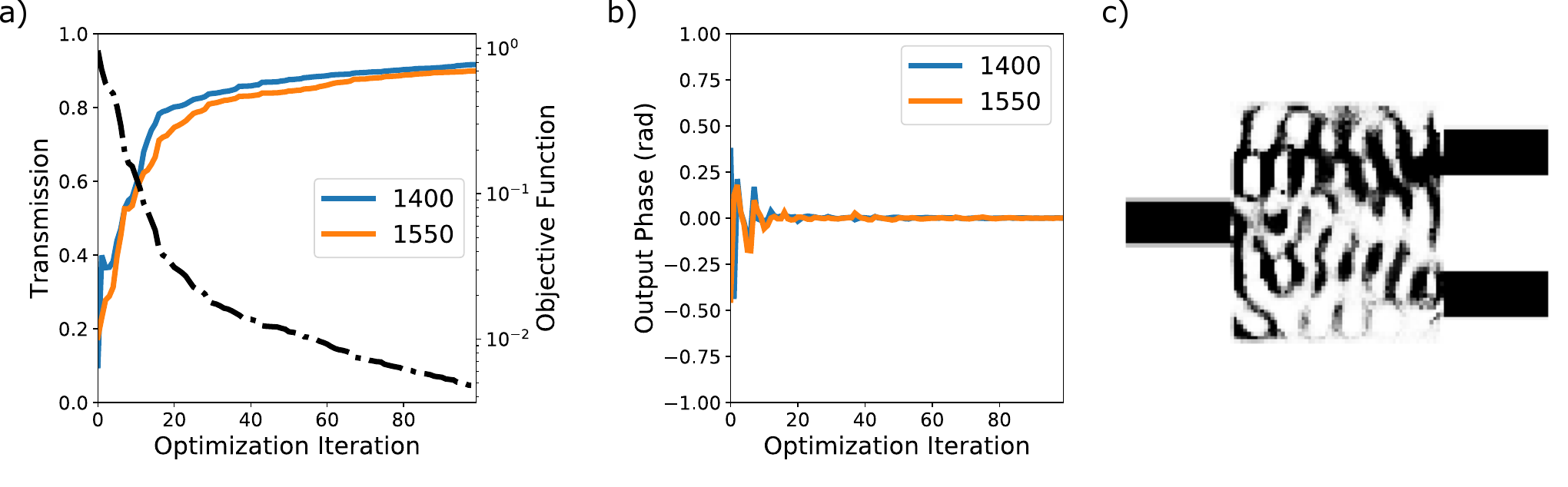}
    \caption{Device B. a) Transmission (solid) and objective value (dotted) as a function of optimization iteration.  b) Phase of the output waveguide mode as a function of optimization iteration. c) Final device design.}
    \label{fig:wdm-device-2}
\end{figure*}
In some applications, the phase of the output is also important. To control the phase, the optimization problem is slightly modified to:
\begin{equation}
\begin{split}
\min_{p,E_{1}, E_{2}} \quad & |c_{t, 1}^\dagger E_{1} - e^{i\theta_1}|^2 + f(|c_{t, 2}^\dagger E_{2}|^2, 0)  \\
& + |c_{b, 2}^\dagger E_{2} - e^{i\theta_2}|^2 + f(|c_{b, 1}^\dagger E_{1}|^2, 0) \\
\textrm{subject to} \quad & \nabla \times \frac{1}{\mu} \nabla \times E_{1} - \omega_1^2 \epsilon(p) E_{1} = -i\omega_{1} J_{1} \\
& \nabla \times \frac{1}{\mu} \nabla \times E_{2} - \omega_2^2 \epsilon(p) E_{2} = -i\omega_{2} J_{2} 
\end{split}
\end{equation}
where $\theta_1$ and $\theta_2$ are the desired phases. In this objective function, the complex overlap value is directly compared against a complex coefficient, so the objective function will be low only if the modal overlap has the right amplitude and phase. 

The parametrization and optimization of this device proceeds similarly to Device 1a (Section \ref{sec:wdm-continuous}) (target phases are set to zero). The final device design is shown in Figure \ref{fig:wdm-device-2}c, and the device achieves 91\% transmission at 1400 and 90\% transmission at 1550 nm with around 19 dB crosstalk suppression. The optimization trajectory is plotted in Figure \ref{fig:wdm-device-2}a, and the trajectory of the phase at the output mode is plotted in Figure \ref{fig:wdm-device-2}b. Without the phase objective, the phase of the output modes differ by nearly 1.5 radians (Figure \ref{fig:wdm-device-1a}b), but with the phase objective, the phase at the output modes both become nearly zero. On the other hand, because of this additional constraint, the device transmission is around 5\% lower. If this performance loss were unacceptable, then the design area could be increased. Because this optimization requires the same number of simulations as in Section \ref{sec:wdm-continuous}, the device also took around 3.5 hours to optimize (on the same hardware), but as before, in a practical setting, the performance of the device can be estimated after running only 20-30 iterations.

\section{Practical Considerations}
\label{sec:analysis}
This section highlights some of the most important considerations when using gradient-based nanophotonic optimization. Addition considerations and heuristics are discussed in Appendix \ref{sec:app-practical-considerations}.

\subsection{Local Minima and Initialization}

Because a local gradient-based optimization algorithm is used, the optimized devices converge to a local mimima with respect to the design parameters. As one might expect, there any many possible local minima. Although converging to the global minima would be ideal, the non-convexity of many electromagnetic design problems means that computationally tractable methods for finding the global minima do not exist. Fortunately, for many design problems, it suffices to find a device that meets the design requirements. As will be demonstrated, when using the continuous relaxation approach presented in Section \ref{sec:overview-transformations}, the local minima encountered have roughly the same performance, so running an optimization several times will often provide a rough estimate of the maximum possible performance of a device.

\begin{figure*}
    \centering
    \includegraphics[width=\textwidth]{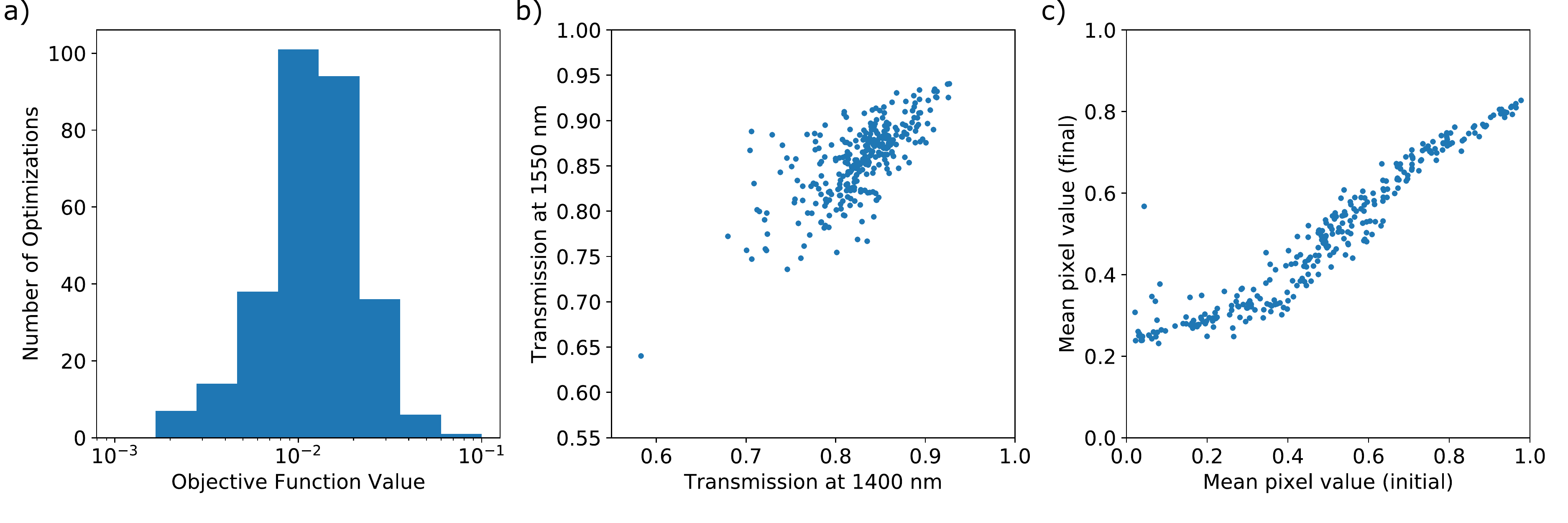}
    \caption{a) Distribution of the objective function values for 297 optimizations. b) The designs show strong correlation between performance at 1400 nm and 1550 nm because of the form of the objective function (see Section \ref{sec:wdm-example-obj}). c) Plot of the mean pixel value of the initial condition (in continuous) versus the mean pixel value for the final structure. There is a strong correlation between the starting pixel value and the final value.}
    \label{fig:wdm-plots}
\end{figure*}

 In particular, we revisit the problem of wavelength-demultiplexing between 1400 nm and 1550 nm presented in Section \ref{sec:wdm-example} and analyze the local minima encountered by repeating the exact same optimization with random initial conditions  (Appendix \ref{sec:app-local-minima}). The distribution of the objective function values and the transmission at 1400 nm and 1550 nm are shown in Figure \ref{fig:wdm-plots}a-b. The objective function distribution demonstrates that local minima encountered have generally similar performance despite having vastly different initial conditions. In particular, most devices have transmission efficiencies between 80\% and 90\% at both 1400 nm and 1550 nm. Considering that a continuous permittivity device achieves around 95\% efficiency (Section \ref{sec:wdm-continuous}), this demonstrates that most of these local minima have decent performance. In fact, only one device has significantly worse performance than the others with around 60\% transmission at both 1400 nm and 1550 nm. Although we are presenting data for the discrete structures, the data for the continuous structures are similar (Appendix \ref{sec:app-local-minima}).
 
 \begin{figure*}
    \centering
    \includegraphics[width=\textwidth]{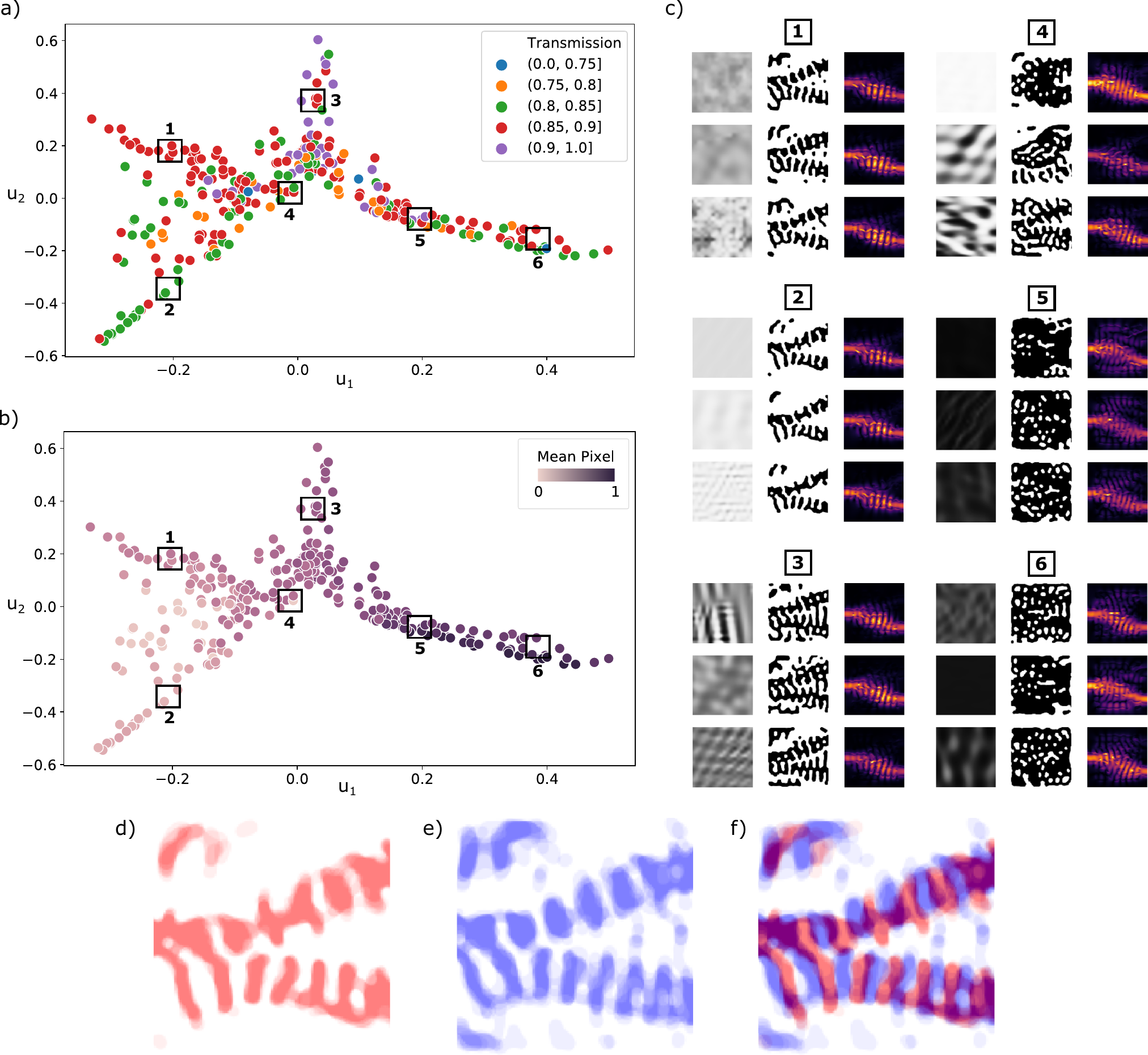}
    \caption{Spectral embedding of the final structures of the optimizations from Figure \ref{fig:wdm-plots}. In a spectral embedding, nearby points represent structures that are visually similar. a) Spectral embedding colored binned by the transmission values at 1550 nm. $u_1$ and $u_2$ correspond to the values of the first and second embedding vector. b) Same spectral embedding as in (a) but colored by the mean pixel value of the initial condition. c) A sample collection of these structures in the spectral embedding. The above each group of three structures correspond to the three closest structures to the region boxed by the corresponding number in the spectral embedding plots. For each group, the left column shows the initial condition, the middle column shows the final structure, and the right column shows the electric field distribution at 1550 nm. d) Average of 8 structures near Box 1. e) Average of 8 structures near Box 2. f) Figures (e) and (f) overlaid on top  of each other. We clearly see that the devices near Box 1 are the "complement" of the devices near Box 2 in the grating region. }
    \label{fig:landscape}
\end{figure*}

Figure \ref{fig:landscape} shows a selection of local minima structures. The minima are organized using a spectral embedding \cite{von2007tutorial} to plot structures in such way that clusters similar structures together (see Supplementary Material). Therefore, nearby points in the 2D plot correspond to structures that resemble each other. The "branches" of the embedding suggest roughly four classes of local minima. First, there are structures that are essentially silicon blocks with various holes (Boxes 5 and 6). Within these structures, the electric field spreads across the entire device, suggesting that multi-path interference contributes to device performance. Second, there are two classes of splitter-like structures (Boxes 1 and 2). In these structures, the electric field is mostly constrained to one of the two paths, and the structures themselves resemble miniature Bragg mirrors that reflect the unwanted wavelength. Interestingly, these two classes differ essentially in the relative placement of the "gratings" (Figure \ref{fig:landscape}d-f), yet despite this similarity, as can be seen by Figure \ref{fig:landscape}a, one class clearly outperforms the other. Third, there is a class of structures that resemble the "splitter-grating" structures but the paths are not quite as separated (Box 3). This class of devices seem to be the best-performing. Finally, the devices in the center of the plot tend to resemble and behave as a mixture of these classes (Box 4).

There is a strong correlation between the mean permittivity of the initial structure and the mean permittivity of the final structure, as shown in Figure \ref{fig:wdm-plots}c. This fact can be used to bias structures towards more silicon or more cladding if it is expected that one regime will perform better than other.  In this case, Figure \ref{fig:landscape} clearly shows that the mean pixel value strongly influences the class of the optimized device. Additionally, notice that the mean pixel value of the final structure always lies roughly between 0.2 and 0.8 regardless of the initial condition, which spans the full range. This arises from the fact that a device with mean value close to 0 is mostly silica and a device with mean value close to 1 is mostly silicon. In both cases, the device performance would be poor, and hence the optimization steers the mean pixel value to within this range.

Even though mean initial pixel value strongly influences the ultimate design, Figure \ref{fig:landscape} demonstrates the value of adding noise and variations on in the initial condition. Even within the same class, performance can have significant variation. This suggests that small variations in the initial condition can lead to local variations in the final structure that affect performance. Consequently, to generate a variety of structures of a certain type, a little bit of Gaussian noise can be added to create slightly different initial conditions. After optimizing each of the devices, the most appropriate device can then be selected from the final structures.

\subsection{Device Performance Bounds}
\label{sec:device-bounds}

\begin{figure}
    \centering
    \includegraphics[width=4in]{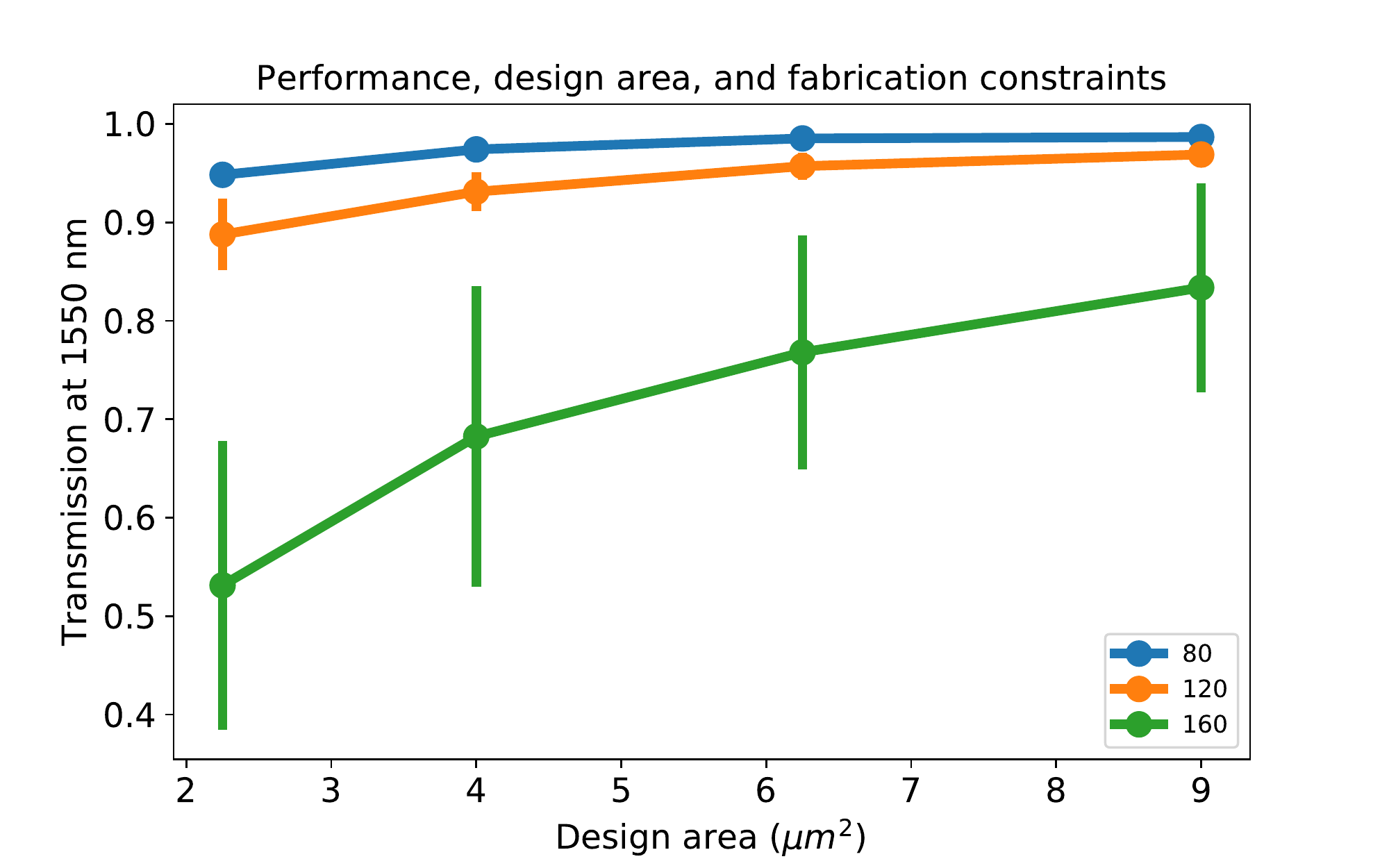}
    \caption{Performance of a simulated device (wavelength demultiplexer in 2D) as a function of design area. The different colors correspond to different feature size constraints. The data is taken from \cite{vercruysse2019analytical}.}
    \label{fig:perf-design-fab}
\end{figure}

Performance of a device is usually affected by fabrication constraints and device design area. In general, increasing the degrees of freedom in an optimization improves device performance. Figure \ref{fig:perf-design-fab} shows how performance is affected for 2D simulations of a wavelength demultiplexer. As feature size approaches zero, the performance should approach the continuous distribution device, since the continuous distributions can be realized by using deep subwavelength features, which, in an effective index approximation, yields the same permittivity. As the feature size becomes larger, worst-performing devices as well as a larger variation in device performance is observed. Likewise, larger design areas results in better performing devices as well as a reduction in the variation in device performance. As a consequence of performance variation, when working with large feature sizes and small design areas, it is important to run more optimizations.

As illustrated in Section \ref{sec:wdm-continuous}, instead of performing full optimizations to estimate the appropriate design area to use, a simple way is to run continuous stage optimization several times for a small number of iterations. This method, however, cannot be used to estimate the performance with a particular feature size. To do so, similar to Section \ref{sec:wdm-discrete}, continuous optimizations using the cubic interpolation parametrization can be run instead.

Sometimes it is not clear that it is physically possible to achieve the desired behavior in a given design region. To date, rigorously bounding performance of devices is only possible in very limited cases and is not necessarily tight \cite{angeris2018computational,shim2019fundamental}. For a rough estimate of an upper bound, one can simply optimize the device without any fabrication constraints, including the top-down lithographic constraint. In other words, the permittivity distribution in the entire volume can be modified and set to intermediate permittivities. This method can also help suggest the ideal fabrication methods required \cite{su2018fully}.

\section{Conclusion}
SPINS enables a flexible nanophotonic design platform built using small building blocks to create an overall optimization plan. SPINS is a design framework, not a methodology. It is a way to structure nanophotonic design problems to make it easy to extend and experiment with them. By relying on backpropagation, gradients are computed automatically, enabling designers to quickly test different objectives. By explicitly laying out dependencies, SPINS can automatically record all the hyperparameters used to define the problem, enabling optimizations to be restarted from any point. Additionally, with explicit dependencies laid out, SPINS opens up the future possibility of optimizing the computational graph itself to maximize performance on the underlying hardware.

As with any complex tool, using inverse design effectively requires understanding the control knobs available to the designer. To that end, the design of wavelength demultiplexers was extensively studied in order to illustrate the practical considerations and heuristics required in using inverse design. These include the choice of objective function, parametrization, and initial condition. In particular, an in-depth analysis of the local minima encountered by the design algorithm provided heuristics for choosing the initial condition. The Appendix includes further guidance on managing electromagnetic simulations, handling discretization, and choosing an appropriate optimizer. By designing a framework with an eye towards these considerations, SPINS aims to enable designers and researchers alike to be more effective in using inverse design.

\section*{Acknowledgements}
We thank Geun Ho Ahn for helping review the manuscript and Rahul Trivedi for helping with implementation of the framework.
D.V. acknowledges funding from FWO and European Union’s Horizon 2020 research and
innovation program under the Marie Sklodowska-Curie grant agreement No 665501. J.S. acknowledges the National Science Foundation Graduate Research Fellowship under Grant No. DGE-165618. We acknowledge funding from Gordon and Betty Moore Foundation, and we thank
Google for providing computational resources on the Google Cloud Platform.

\section*{Competing Interests}
All authors receive royalties from the licensed version of SPINS.

\bibliographystyle{unsrt}
\bibliography{article}

\clearpage
\appendix

\section{Framework Details}
\label{sec:app-frame-details}

This section explains the details of the framework. The subsections titled "Implementation Details" contain deeper implementation details that may be skipped without a loss of understanding.

\subsection{Parametrization, Selection Matrix, and Permittivity Distribution}
\label{sec:selection-matrix}
As discussed in Section \ref{sec:overview-parametrization}, it is usually beneficial to describe the permittivity distribution via a parametrization that better captures the true degrees of freedom in the design region. For many devices, the top-down lithographic constraint means that a 2D image is sufficient to describe the full 3D permittivity distribution. This motivates the introduction of a {\it selection matrix} defined through the equation:
\begin{align}
    \epsilon(p) = \epsilon_{bg} + S\theta(p)
\end{align}
where $\epsilon$ is the permittivity distribution, $\epsilon_{bg}$ is a constant permittivity background, and $\theta$ ranges from 0 to 1. Often, $\theta(p)$ represents a 2D slice of the permittivity distribution, and therefore the dimensions of $\theta$ are usually much less than the dimensions of $\epsilon$. However, there is not a loss of generality as $S$ can always be taken to be the identity matrix.

Defining the selection matrix has several additional advantages. First, the selection matrix can capture any additional equality constraints on the permittivity distribution, such as structure symmetry and periodicity. Second, the selection matrix is scaled so that $\theta$ is normalized between 0 and 1. This normalization allows parametrizations to be agnostic to the specific device materials and hence makes parametrizations more general and flexible. Last, the selection matrix conveniently handles the details of permittivity averaging on the simulation grid (Appendix \ref{sec:app-sim-impl-details}).

As a consequence of this definition, the parametrization is technically a function that maps a parametrization vector $p$ into $\theta$. As mentioned in Section \ref{sec:overview-parametrization}, there are many approaches to designing parametrizations. The simplest parametrization is the direct parametrization that assigns $\theta(p) = p$. Though this may not be ideal for imposing fabrication constraints, it is a useful parametrization for determining the required device area (see Section \ref{sec:device-bounds}). Examples of other possible parametrizations include one that defines a structure based on the boundary of a device \cite{michaels2018leveraging} and one that defines the structure through a set of rectangles that have different positions and sizes \cite{jan2016thesis}. In general, parametrizations should have large degrees of freedom in order to not unnecessarily restrict the design space. Note that for low-dimensional parametrizations, Bayesian optimization and even brute force parameter sweeps may be a better choice than inverse design \cite{schneider2018benchmarking}.

\subsubsection{Levelsets}
\label{sec:levelsets}
\begin{figure}
    \centering
    \includegraphics[width=\columnwidth]{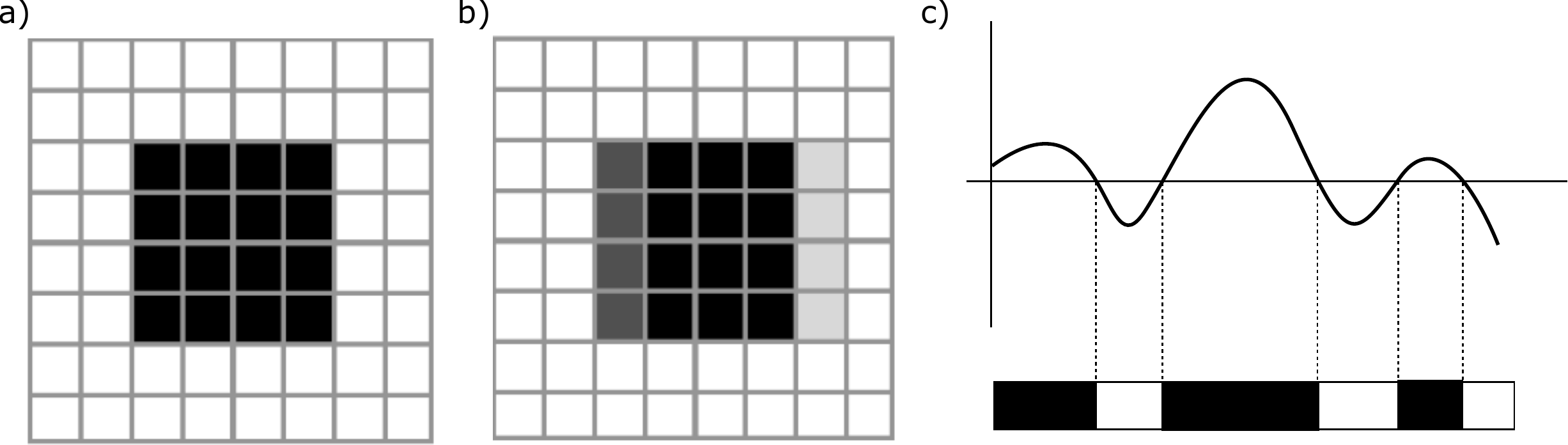}
    \caption{a-b) When the device is perfectly aligned to the grid (a), then binary pixels may be possible, but when the same device is misaligned (b), the binary device results in non-binary pixels. c) A binary device is represented by a levelset function using the following rule: When the levelset function is above zero, one material is used, and when the levelset function is below zero, another material is used. In other words, the zero-contour of the levelset function forms the contour of the device. Notice that by construction, the levelset function always defines a binary device.}
    \label{fig:levelsets}
\end{figure}

For lithographically-defined devices in discrete optimization, levelset parametrizations are convenient because they always define a discrete structure. In the levelset parametrization, a binary permittivity distribution is defined via a levelset function. Whenever the function is above zero, the permittivity has one value, and whenever the function is below zero, the permittivity has another value. In other words, the contour of the device is defined by where the function crosses zero (Figure \ref{fig:levelsets}b).

The advantage of levelsets over the direct parametrization is two-fold. First, binary devices need not correspond to binary pixels (Figure \ref{fig:levelsets}a), so simply enforcing that individual pixels be binary is actually over-constraining the design space. Second, defining differentiable fabrication constraints such as minimum feature size is challenging with the direct parametrization whereas this is more natural for levelsets \cite{vercruysse2019analytical}.

\subsubsection{Implementation Details}
Note that the selection matrix defined this way can handle arbitrarily-shaped polygons. For example, this can be used to ensure a gap with the design region where there is no structure \cite{sapra2019chip}. Consequently, as the selection matrix can be complicated to specify by hand, SPINS generates the selection using {\it foreground} and {\it background} permittivity distributions. The foreground distribution corresponds to the permittivity when $\theta$ is all ones whereas the background distribution corresponds to the permittivity when $\theta$ is all zeros. The selection matrix can then be generated in two steps. First, a foreground and background permittivity distribution are subtracted to obtain a mask indicating where the permittivity is permitted to change (i.e. anywhere that is non-zero). Based on this mask, a uniform selection matrix is constructed by sampling twice as finely as the Yee grid over the design area. The permittivity values on the Yee grid are then determined through appropriate averaging over permittivity values defined on this finer grid. Second, this uniform selection matrix is weighted (i.e. multiplied element-wise) by this mask. Last, any symmetry conditions are applied.

\subsection{Simulation}

From the perspective of the computational graph, the simulation node accepts the permittivity distribution and produces the corresponding fields. However, other quantities, including sources and boundary conditions, must also be specified. In SPINS, sources and the {\it simulation space}, an object that describes both the simulation region and boundary conditions, are themselves problem graph nodes. This way, the actual simulation node is focused solely on running the simulation and is agnostic to specific sources or simulation spaces.

An appropriate simulation method should be chosen for the simulation node because optimization time is dominated by the simulation time. For example, for waveguide-based devices where performance only matters at a few wavelengths, it is generally faster to use finite-difference frequency-domain (FDFD) method over finite-difference time-domain (FDTD) method. However, for devices requiring optimization over a broad spectrum of wavelengths, FDTD may be more appropriate. For devices with periodic boundary conditions, such as in metasurface design, rigorous coupled wave analysis (RCWA) may be faster. In this manuscript, we focus on FDFD because that is the only implemented solver in SPINS so far. However, SPINS as a framework is agnostic to the exact simulation method used.

\subsubsection{Implementation Details}
\label{sec:app-sim-impl-details}
SPINS maintains its own implementation of the FDFD algorithm, including its own mode solvers. This is necessary as accurate implementation of adjoint calculations require the ability to precisely specify the problem to the solver (e.g. potentially requires arbitrary specification of sources). Radiation boundary conditions are handled using SC-PMLs \cite{Shin12}.

In general, 2D simulation can be treated as a special case of a 3D simulation with perfect translation symmetry along one axis. This is implemented as an axis with exactly one FDFD cell and periodic boundary conditions. Such an implementation obviates the need to special case code depending on the dimensionality of the simulation at the cost of some unnecessary memory usage for 2D simulations. However, different types of matrix solvers are used depending on the dimensionality because of the different memory requirements.

For 2D simulations, direct matrix inversion is usually possible and empirically observed to be faster than iterative methods. Moreover, since the solve time for 2D simulations is fairly quick, SPINS by default runs 2D simulations directly on the CPU. This operation can be done directly on CPUs using standard linear algebra libraries.

For 3D simulations, inverting the Maxwell operator directly becomes computationally intractable from both a time and memory standpoint. As a consequence, an iterative solver is used instead in which only matrix-vector products are required. When SC-PMLs are applied with periodic boundary conditions, a symmetrizer \cite{Shin12, Shin13} can be applied in order to use COCG \cite{van1990petrov}. Note that with PMLs and the symmetrizer, the Maxwell operator is {\it complex symmetric} rather than Hermitian, and therefore many convergence guarantees associated with numerical methods do not apply. Nevertheless from \cite{erlangga2008advance}, we see that COCG typically fairs well, though introduction of Bloch boundary conditions requires a non-symmetric solver (e.g. BiCG-STAB).

Simulation time is the main bottleneck in electromagnetic optimization. Consequently, SPINS uses a multi-GPU implementation of Maxwell equations. To use multiple GPUs, the simulation region is sliced along one simulation axis, and then each region is assigned to one GPU. However, the number of GPUs that can exist on a system is limited. Note that using GPUs located physically on different motherboards is undesirable because GPUs must share information every iteration. However, such systems are still useful for optimizations that require multiple independent simulations (e.g. for multiple wavelengths).

\subsection{Objectives and Function Nodes}
The objective function is described by a set of generic function nodes. Function nodes accept one or more input arguments and generally produce a single output. Functions also can compute the gradient with respect to any of its inputs. These function nodes range from simple operations, such as addition and exponentiation, to more complicated nodes, such as log-sum-exp. Generally speaking, simple nodes are preferred as they allow for more flexible composability. However, in some cases, complicated operations are implemented as a single node in order to optimize the performance. For example, log-sum-exp is a single node because a naive implementation could easily result in overflow/underflow errors. Likewise, the FDFD simulation node could technically be broken down into smaller operations, but this would be at the cost of operation speed.

\subsection{Transformations}
The optimization is executed through a series of {\it transformations}. In the most general sense, a transformation is simply an operation that affects the state of optimization. Transformations can be complex operations such as finding the parametrization that minimizes an objective function or smaller operations such as performing thresholding on a structure. In general, the full optimization in SPINS consists of several sub-optimizations, each of which is implemented as a transformation. Note that the continuous and discrete optimization stages described in Section \ref{sec:overview-transformations} may actually consist of several transformations. For example, the continuous stage may consist of multiple transformations that act on structures that become increasingly more discrete in nature \cite{vercruysse2019analytical}. The use of transformations enables different parametrizations and optimization stages to be easily swapped in and out for one another. 

Many transformations accept an {\it optimizer} as an argument. An optimizer implements a specific optimization method to solve a generic (i.e. unrelated to electromagnetics) problem. SPINS defines a lightweight interface for optimizers so that optimizers from different existing third-party implementations can be used. For example, SPINS users commonly rely on the SciPy optimizer which wraps the optimization functions implemented in SciPy. Higher-order optimizers, i.e. those that rely on other optimizers, can also be implemented. For instance, SPINS provides an optimizer to solve constrained optimization problem using unconstrained optimizers through the augmented Lagrangian formalism.

\subsection{Executing the Optimization Plan}
To actually perform the optimization, the transformations are executed in order, one after another. During the execution of a transformation, the transformation will reference nodes in the graph, such as the node representing the objective function. Whenever this happens, that node will be materialized into a Python object through the {\it workspace}. In other words, the workspace accepts a node and returns an object that can be used to perform actual calculations. Since the problem graph contains information about all the dependencies associated with a given node, the workspace will materialize only the nodes necessary, thus avoiding any unnecessary compute.

Because there is a one-to-one mapping between a problem graph node and a Python object, a computational graph that mirrors the function nodes in the problem graph can be constructed. This computational graph is used to both evaluate functions and compute their gradients. Because the computational graph contains all the explicit dependencies between operations, the computations can be automatically parallelized. In practice, the most time-consuming operation is the electromagnetic simulations. Therefore, the electromagnetic simulations are executed in parallel as much as possible, whereas the remaining computations are performed serially.

The gradient is computed using reverse-mode autodifferentiation, also commonly known as backpropagation, which is an efficient method to compute the gradient of a function with few outputs (in this case one) and many inputs (dimensionality of the parametrization). Backpropagation through the electromagnetic simulation is also often referred to as the adjoint method or adjoint simulation \cite{lee1997systematic}. The special name arises from the fact that the computation involves another electromagnetic simulation with the same permittivity but different source. 

From an implementation perspective, SPINS currently chooses to implement its own backpropagation for two reasons. First, backpropagation through complex-valued functions is required and must be handled appropriately, but many existing libraries do not fully support or have proper complex-valued backpropagation. Second, the backpropagation code must be able to properly handle parallelization of electromagnetic simulations.

\subsubsection{Monitoring and Logging}
One advantage of using a graph is that monitoring and logging become straightforward. Monitors are objects that capture the output of a particular node in the computational graph and save the data into log files. This enables probing any value used during the optimization. For example, the electric field values can be saved by probing the simulation node, and any sub-objectives can be saved by probing the corresponding function nodes. Additional nodes that are not used in the optimization can be added and monitored to provide more information. For instance, a mode monitor could be added at the input port to monitor back-reflection even if the objective function does not depend on back-reflection.

SPINS enforces a standard logging format that stores the full state of the optimization at any given time. In this way, post-optimization analysis tools can be shared. Moreover, since the optimization plan is an exhaustive description of the design problem, SPINS is able to restore any optimization from any iteration by checking the state of the optimization in the log files. Not only is this feature invaluable for restarting optimizations that fail part-way, but it is also invaluable for avoiding re-execution of transformations when experimenting with transformation sequences. For instance, to run a different discrete optimization transformation, SPINS can simply restore the optimization up to the discretization transformation and then execute the new discrete transformation, without having to re-run any of the previous continuous transformations.

\section{More Practical Considerations}
\label{sec:app-practical-considerations}
\subsection{FDFD Simulation Considerations}
\label{sec:grid-spacing}
Optimization of nanophotonic devices, particularly at the beginning of the optimization, do not require highly accurate simulations. As a result, computation time can be reduced by running lower quality simulations than is normally done for general electromagnetic simulations.

The grid spacing should be maximized as much as possible to reduce the computational load. Empirically, we have found that $\Delta x \approx \lambda / 10$ where $\lambda$ is the wavelength of light in the material is sufficient during optimization. Moreover, we often find that in the absence of highly precise fabrication, using significantly denser grid spacings to optimize is not particularly helpful because the fabrication errors result in much larger performance degradation than compared to errors in the simulation. However, this coarse discretization is only possible because of permittivity averaging.

Permittivity averaging at the boundary between two dielectric materials is critical to having accurate enough simulations at relatively large grid spacings. In SPINS, the permittivity on the border of two materials is implemented as the weighted average of the permittivities based on the fraction of the cell that the materials occupy. Although there are more sophisticated permittivity averaging methods \cite{farjadpour2006improving}, they require simulating with a full permittivity tensor. Note that because of the staggered nature of the Yee grid, for 3D devices, the permittivity values require averaging even if the pixels are perfectly binary.

In addition to the spacing, the error threshold for the solver can be set higher than what would normally be desirable in a highly accurate simulation. In addition, the fields only need to be accurate where the fields are actually used in the computation, which, depending on the solver used, can occur at unexpected thresholds. For example, when using COCG iterative method, the fields in the PMLs take the longest to converge, but they are also not relevant in the simulation. Note that if accurate simulations are required (e.g. in design of a resonator), it is still possible to optimize early on with rough simulations and then switch to finer resolution simulation later on.

\begin{figure}
    \centering
    \includegraphics[width=\columnwidth]{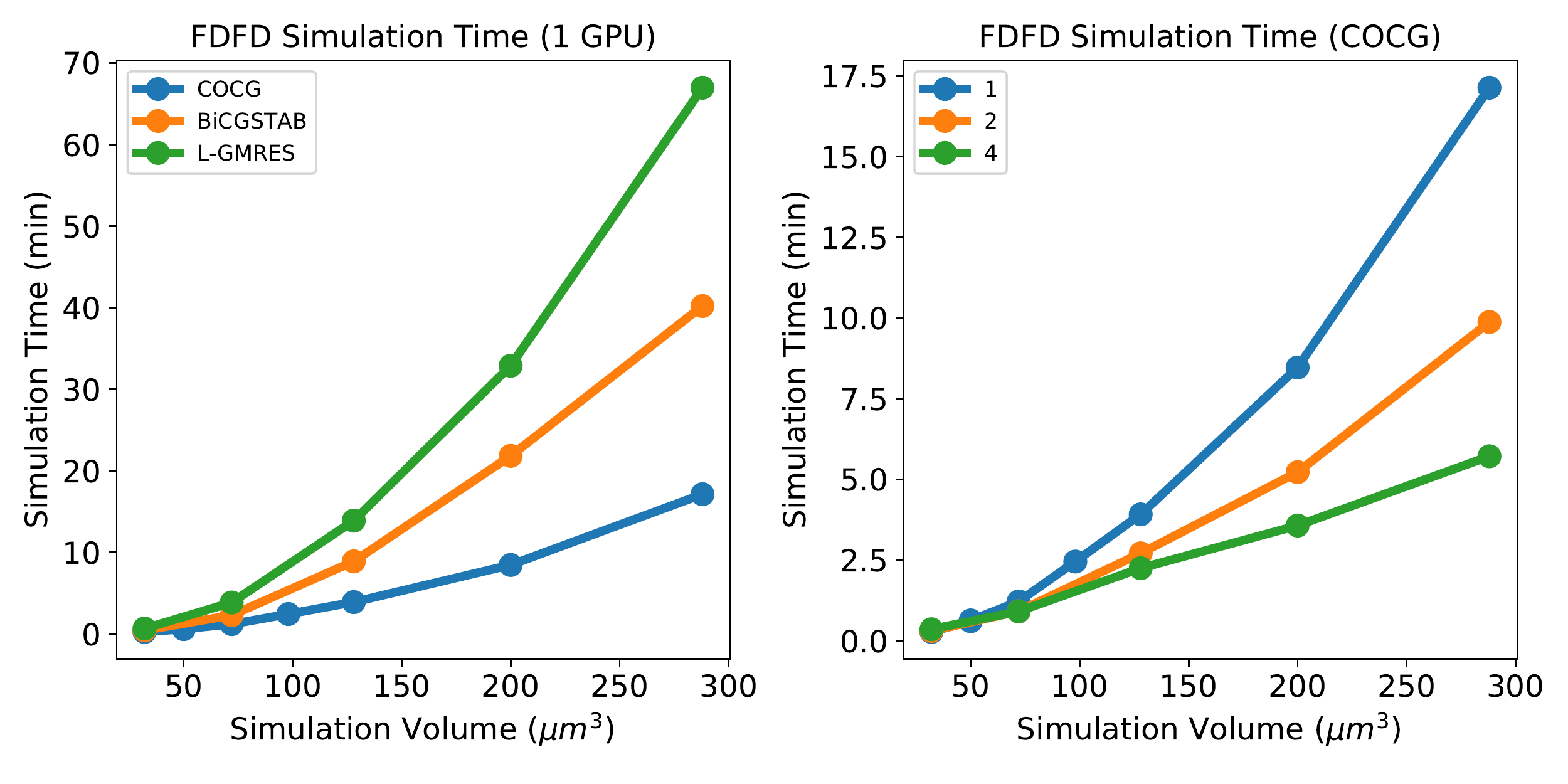}
    \caption{FDFD simulations on NVIDIA K80s. a) FDFD simulation time as a function of simulation volume for different numerical methods to achieve $10^{-6}$ error. b) FDFD simulation time for GPU-accelerated FDFD simulations with different numbers of GPUs.}
    \label{fig:fdfd-benchmarks}
\end{figure}

Beyond approximating simulations, it is equally important to ensure to use the appropriate matrix solvers to run the simulation. For instance, empirically the matrix solver UMFPACK \cite{davis2004algorithm} performs substantially faster than the default matrix solver SuperLU \cite{li2005overview} provided in the SciPy package for 2D electromagnetic simulations. For multi-GPU accelerated FDFD simulations, it is important to pick the correct number of GPUs used per simulation as using the maximum number of GPUs available per simulation may not actually be faster (Figure \ref{fig:fdfd-benchmarks}). This is particularly true for multi-wavelength optimizations where multiple simulations can run simultaneously. For example, for a system with 2 GPUs and an optimization requiring 2 EM simulations, it is likely faster to run each simulation on 1 GPU than to run each simulation on 2 GPUs one after another.

\subsection{Discretization}
\label{sec:discretization}
As the continuous optimization can be thought of as complex initialization procedure for discrete optimization, it is important that the discretization process does not destroy the progress made by the continuous optimization. Empirically, the discretization procedure rarely maintains the performance of the device and often causes the performance of the device to drop significantly. Therefore, a good discretization procedure is one that allows the optimization to eventually recover the performance encountered in the continuous optimization.

Empirically, two main factors that affect discretization are ``discreteness" and ``feature sizes" of the continuous structure. Discreteness is the extent to which the structure already has a binary permittivity. Unsurprisingly, continuous structures that are mostly discrete tend to perform better than structures that have intermediate permittivities. Sometimes the optimization process naturally generates a mostly discrete device. However, this is not guaranteed, and running optimization for longer does not necessarily generate a fully discrete structure. In other words, there exist local minima in which permittivity values are not discrete. Mitigation methods include modifying the continuous optimization to incorporate a discreteness penalty \cite{su2017inverse} or forcing the continuous structure to be more discrete \cite{vercruysse2019analytical}.

Likewise, discretization of a device with large feature size constraints tends to be more difficult. Of course, the notion of feature size is not well-defined in a continuous structure, but as a proxy, the feature size of the thresholded structure can be considered as an estimate of the feature size of the continuous structure. Discretization tends to do poorly if the ``feature size" of the continuous structure is much smaller than the feature size of the discrete structure.

In order to facilitate better discretization, a variety of techniques have been incorporated into the continuous stage optimizations. In order to address the feature constraint issue, \cite{vercruysse2019analytical} introduced a cubically-interpolated continuous parametrization so that the ``feature size" of the continuous structure is well-mapped to the feature size of the discrete structure. In order to mitigate the ``discreteness" issue, several continuous optimizations are run, each with increasing discreteness in the structure. In contrast, \cite{zhou2015minimum} uses filtered versions of a direct parametrization to impose feature size constraints on continuous structures. We refer the interested reader to the literature in topology optimization for other options \cite{sigmund2009manufacturing, wang2011projection}.

\subsection{Crafting Objective Functions}
\label{sec:objective-funs}
As discussed previously, the objective function is an important manner in which the designer can affect the performance of an optimization. Section \ref{sec:wdm-example} already illustrated a common objective function for maximizing transmission through a given port. This section goes into more depth and illustrates other common forms for the objective function in electromagnetic optimization. We refer the reader to a vast body of inverse design literature for inspiration on additional forms of objectives functions \cite{sapra2019chip,hughes2018adjoint,liang2013formulation,lu2013nanophotonic,lin2016cavity,sell2017large}.

\subsubsection{Phase Objective}
Transmission and phase can be simultaneously optimized with an objective of the form $f(b) = |b - b_t|^2$ where $b$ is the complex scattering matrix element for an output port and $b_t = a_t \exp(i \theta_t)$ is the target matrix element with amplitude $a_t$ and phase $\theta_t$. As transmission and phase correspond to two degrees of freedom, this objective implicitly defines a trade-off between transmission and phase. For example, if the phase is far from $\theta_t$ but the amplitude is close to $a_t$, then the optimization will focus more on adjusting the phase than changing the amplitude. If both amplitude and phase are far from the targets, then depending on the actual values, the optimization will either favor adjusting amplitude or adjusting phase. Sometimes this is problematic if $a_t$ is set to unrealistic values (e.g. unity transmission for a device with too small of a design region) as the optimization may keep trying to optimize solely for amplitude while neglecting phase.

\subsubsection{Aggregating Sub-objectives}
As illustrated in Section \ref{sec:wdm-example}, electromagnetic design problems are often multi-objective optimization problems and consist of many different sub-objectives. However, these sub-objectives must be combined into a single total objective function, the construction of which implicitly defines the trade-off between the sub-objectives. Therefore, depending on the desired performance characteristics, one may want to modify how the sub-objectives are combined.

Optimization theory provides a systematic way to trace out the trade-off curve by using an objective that sums the sub-objectives together with variable weights. By setting the weights accordingly, the entire trade-off curve can be determined \cite{boyd2004convex}. Unfortunately, it is often difficult to choose the appropriate weights a priori and thus determining the weights is often a matter of trial-and-error.

In many cases, electromagnetic designers want to give roughly equal weight to the sub-objectives. In these cases, an objective of the form $f(p) = \max f_i(p)$ can be used as an alternative. This forces the optimization to improve the worst-performing metric. Sometimes this optimization works better by using a smooth approximation of the maximum function, e.g. using log-sum-exp or an objective of the form $f(p) = \sum_i f_i(p)^q$ for some integer power $q$.

\subsubsection{Broadband Objective}
\label{sec:objective-funs-broadband}
Devices can be made more broadband by adding sub-objectives at nearby wavelengths. For example, suppose the objective is given by $f_\lambda (p)$ at wavelength $\lambda$, then the corresponding broadband objective would be $f(p) = \sum_{i=-n}^{n} f_{\lambda + i\Delta}(p)$. Broadband performance across a given bandwidth can be achieved by picking $n$ and $\Delta$ judiciously. Usually, this hinges at choosing an appropriate value for $\Delta$. If it is too big, then the spectra may have undesired behavior in between the simulated wavelengths, undermining the broadband performance objective. If it is too small, then $n$ is forced to be large to cover a larger range, and consequently the optimization takes more time. It is recommended to run some quick continuous optimizations (see Section \ref{sec:device-bounds}) to determine what an appropriate value of $\Delta$ should be.

\subsection{Selecting an Optimizer}

The optimizer is responsible for accepting an optimization problem and minimizing the objective function with respect to any constraints. There are a wide variety of optimizations methods that can be used, and we refer the reader to \cite{nocedal2006numerical} for details. Optimization methods can be broadly classified based on whether they employ the gradient to optimize a device. Gradient-free methods, including particle swarm optimization, genetic optimization, Nelder-Mead, and Bayesian optimization, try to sample the objective function at cleverly chosen points based on previous observations. Nevertheless, they typically require many function evaluations as the dimension of the design space increases and are typically ineffective in high dimensional design spaces in which SPINS is designed to operate. 

For unconstrained optimization, gradient-based optimization methods can be broadly classified as first-order, second-order, and quasi-Newton methods depending on how gradient information is used. In first-order methods, only the gradient (as opposed to the Hessian) is used. These include vanilla gradient descent as well as many methods employed in machine learning, such as Adagrad, ADAM, and RMSProp. First-order methods generally take longer to converge as compared to higher-order methods and require setting an appropriate scaling of the step size parameter to achieve convergence, hence requiring hyperparameter tuning to use. In contrast to ML loss functions, evaluating EM objective functions is computationally expensive and should be avoided.

On the other hand, second-order methods, such as Newton's method, that take into account the Hessian of the objective function would be ideal. Because second-order methods have information about the local curvature, they generally converge faster than first-order methods. Unfortunately, it is computationally intractable to actually compute the Hessian for EM optimizations. Instead, we rely on quasi-Newton methods that approximate the Hessian by storing previous gradients. In practice, we default to the SciPy implementation of L-BFGS-B, which empirically works well relative to first-order methods and other optimization methods implemented by SciPy. Additionally, it is empirically observed that L-BFGS-B tends to create structures that are more discrete in the continuous stage optimization, leading to better discretization.

Nevertheless, the best optimization method is yet another knob of control that the designer can use to improve designs.
For cavity designs, for instance, it is observed that using a combination of MMA and L-BFGS-B works better. Since L-BFGS-B cannot handle arbitrary constraints, SLSQP was used in grating optimization. To date, interior point optimization methods have not been applied to inverse design, and there is possibility that they can outperform L-BFGS-B.

\section{Mathematical Details}
\label{sec:app-math-details}
A general electromagnetic design problem can be expressed in the form:
\begin{equation}
	\begin{split}
	\min_{p} \quad & f_{obj}(E_1, E_2, \dots, E_n, \epsilon_1, \epsilon_2, \dots, \epsilon_n, p)  \\
	\textrm{subject to} \quad &  g_i(p) = 0 \quad\quad i = 1, \dots, m \\
	&  h_i(p) \leq 0 \quad\quad i = 1, \dots, q \\
	\end{split}
\end{equation}

where $E_i$ is the electric field corresponding to the structure $\epsilon_i$ (which depend on $p$). Note that there is significant flexibility in what each $\epsilon_i$ and $E_i$ correspond to. $E_i$ could represent frequency-domain fields produced by FDFD or time-domain fields produced by FDTD. Each $\epsilon_i$ can correspond to the permittivity at a different wavelength, temperature, or carrier concentration (e.g. for active device design). They could also correspond enlarged and shrunken versions of the structure to model fabrication errors. The $g_i(p)$ and $h_i(p)$ terms capture any desired constraints on $p$, particularly fabrication constraints.

As discussed in more detail in Section \ref{sec:selection-matrix}, it is often convenient to define $\epsilon$ through selection matrices $S_i$:
\begin{align}
    \epsilon_i = \epsilon_{0,i} + S_i\theta_i(p)
\end{align}
The function $\theta_i$ is known as the {\it parametrization} and therefore $p$ is the {\it parametrization vector}. Since the output dimension of a parametrization depends on the shape of the selection matrix $S_i$, parametrizations are often designed to work for a particular type of selection matrix. 

To perform gradient-based optimization, the gradient $df_{obj}/dp$ must be computed. Caution must be taken if $E_i$ or $\epsilon_i$ is complex-valued. Since $f_{obj}$ is necessarily non-holomorphic (it maps a complex value to a real value), the complex derivative does not exist. In general,
\begin{align}
   \label{eqn:deriv}
    \frac{df_{obj}}{dp} = \frac{\partial f_{obj}}{\partial p} + \sum_i \Bigg(& \frac{\partial f_{obj}}{\partial E_i}\frac{d E_i}{dp} + \frac{\partial f_{obj}}{\partial E_i^*}\frac{dE_i^*}{dp} 
    + \frac{\partial f_{obj}}{\partial \epsilon_i}\frac{d\epsilon_i}{dp} + \frac{\partial f_{obj}}{\partial \epsilon_i^*}\frac{d\epsilon_i^*}{dp}\Bigg)
\end{align}
where the partial derivatives are defined as Wirtinger derivatives. For a function $f(z)$ where $z = x + iy$, the Wirtinger derivatives are defined as:
\begin{align}
    \frac{\partial f}{\partial z} &= \frac{1}{2}\left(\frac{\partial f}{\partial x} - i\frac{\partial f}{\partial y} \right) \\
    \frac{\partial f}{\partial z^*} &= \frac{1}{2}\left(\frac{\partial f}{\partial x} + i\frac{\partial f}{\partial y} \right)
\end{align}
If $f(z)$ is holomorphic (e.g. $f(z) = z^2$), the Wirtinger derivative $\partial f/\partial z$ is equivalent to the complex derivative $df/dz$. However, the Wirtinger derivative is still well-defined even if $f(z)$ is non-holomorphic (e.g. $f(z) = |z|^2$). If $f(z)$ is real, then $\partial f/\partial z = (\partial f/\partial z^*)^*$. Since $f_{obj}$ is real-valued, Equation \ref{eqn:deriv} simplifies to
\begin{align}
        \frac{df_{obj}}{dp} = \frac{\partial f_{obj}}{\partial p} + 2\operatorname{Re}\Bigg[ \sum_i \Bigg(& \frac{\partial f_{obj}}{\partial E_i}\frac{dE_i}{dp}
    + \frac{\partial f_{obj}}{\partial \epsilon_i}\frac{d\epsilon_i}{dp}\Bigg)\Bigg]
\end{align}
The computation for $\partial f_{obj}/\partial p$, $\partial f_{obj}/\partial E_i$, and $\partial f_{obj}/\partial \epsilon_i$ depends on the form of the objective. The structure gradient $d\epsilon_i/dp$ is straightforward:
\begin{align}
    \frac{d\epsilon_i}{dp} &= \frac{d\epsilon_i}{d\theta_i}\frac{d\theta_i}{dp} \\
    &= S_i\frac{d\theta_i}{dp}
\end{align}

To derive the simulation gradient $dE_i/dp$, we must differentiate through the electromagnetic simulation. We will derive the gradient for FDFD, but similar derivations follow for other computational methods. The FDFD equation is given by
\begin{align}
    (D - \omega^2\text{diag}(\epsilon))E = -i\omega J
\end{align}
where $D$ is the discretized version of the $\nabla\times\frac{1}{\mu}\nabla\times$ operator (permeability is assumed to be constant). Differentiating by through by $\epsilon$ and rearranging, we have that
\begin{align}
    (D - \omega^2\text{diag}(\epsilon))\frac{dE}{d\epsilon} = \omega^2\text{diag}(E)
\end{align} Therefore,
\begin{align}
    \frac{dE_i}{dp} &= \frac{dE_i}{d\epsilon_i}\frac{d\epsilon_i}{dp} \\
    &= (D - \omega_i^2\text{diag}(\epsilon_i))^{-1}\omega_i^2\text{diag}(E_i)\frac{d\epsilon_i}{dp}
\end{align}
Note, however, that computing $dE_i/dp$ is computationally expensive as this requires a number of electromagnetic simulations equal to the number of elements in $p$. Instead, to evaluate the gradient, we rely on backpropagation to reduce the number of simulations to one. During backpropagation, $\partial f_{obj}/\partial E_i$ is computed first and then the quantity $(\partial f_{obj}/\partial E_i)(dE_i/dp)$ is evaluated as:
\begin{align}
    \frac{\partial f_{obj}}{\partial E_i}\frac{dE_i}{dp}
    &= \frac{\partial f_{obj}}{\partial E_i}(D - \omega_i^2\text{diag}(\epsilon_i))^{-1}\omega_i^2\text{diag}(E_i)\frac{d\epsilon_i}{dp} \\
    &= \left((D - \omega_i^2\text{diag}(\epsilon_i))^{-T}\frac{\partial f_{obj}}{\partial E_i}^T\right)^T\omega_i^2\text{diag}(E_i)\frac{d\epsilon_i}{dp}
\end{align}
Consequently, differentiating through FDFD involves an electromagnetic simulation with $(\partial f_{obj}/\partial E_i)^T/(-i\omega_i)$ as the source. Similar results exist for other simulation methods \cite{lee1997systematic,nikolova2004sensitivity,sell2017large}.

\section{Local Minima Analysis Details}
\label{sec:app-local-minima}
In order to thoroughly explore the space of possible local minima, three different initialization methods were used to generate the structures: blurred random noise, Perlin noise, and Gabor noise. The use of different initialization methods is to ensure that most relevant types of local minima are captured in the process. For the blurred random noise structures, the parametrization values are generated uniformly at random and then filtered with a Gaussian kernel of random width. A randomly generated zoom and rotation was then applied. Perlin and Gabor noise are common noise types used in procedural computer graphics. Perlin noise can generate noise with different feature sizes and turbulence, whereas Gabor noise can generate noise with different frequency components. To ensure that the starting conditions had varied amplitudes and average permittivity values, the Perlin and Gabor noise structures were post-processed with a random scaling and random offset. No correlation was observed between the initialization method and the final performance, suggesting that the choice of random initialization may not be particularly significant.

Because the structures live in a high-dimensional space, an embedding method, which maps vectors from a high-dimensional space to a low-dimensional space, must be used to visualize the space of a devices. Principal component analysis (PCA) is perhaps the most well-known embedding methodology, but by virtue of its linearity, may not necessarily be the most effective way to visualize the space if it is nonlinear. In contrast, spectral embedding is a nonlinear embedding that derives from spectral decomposition of undirected graphs and is commonly used to separate a graph into clusters based on the connectivity. For example, if a graph has two connected components, then the embedding vectors will identify which nodes belong to which component. The graph used for spectral embedding is one formed by connecting each structure with the $n$ nearest neighbors where the distance between two structures is defined using the 2-norm. In this analysis, the spectral embedding was performed by scikit-learn \cite{scikit-learn} with $n=10$ nearest neighbors and default values for the other parameters.

\begin{figure*}
    \centering
    \includegraphics[width=4in]{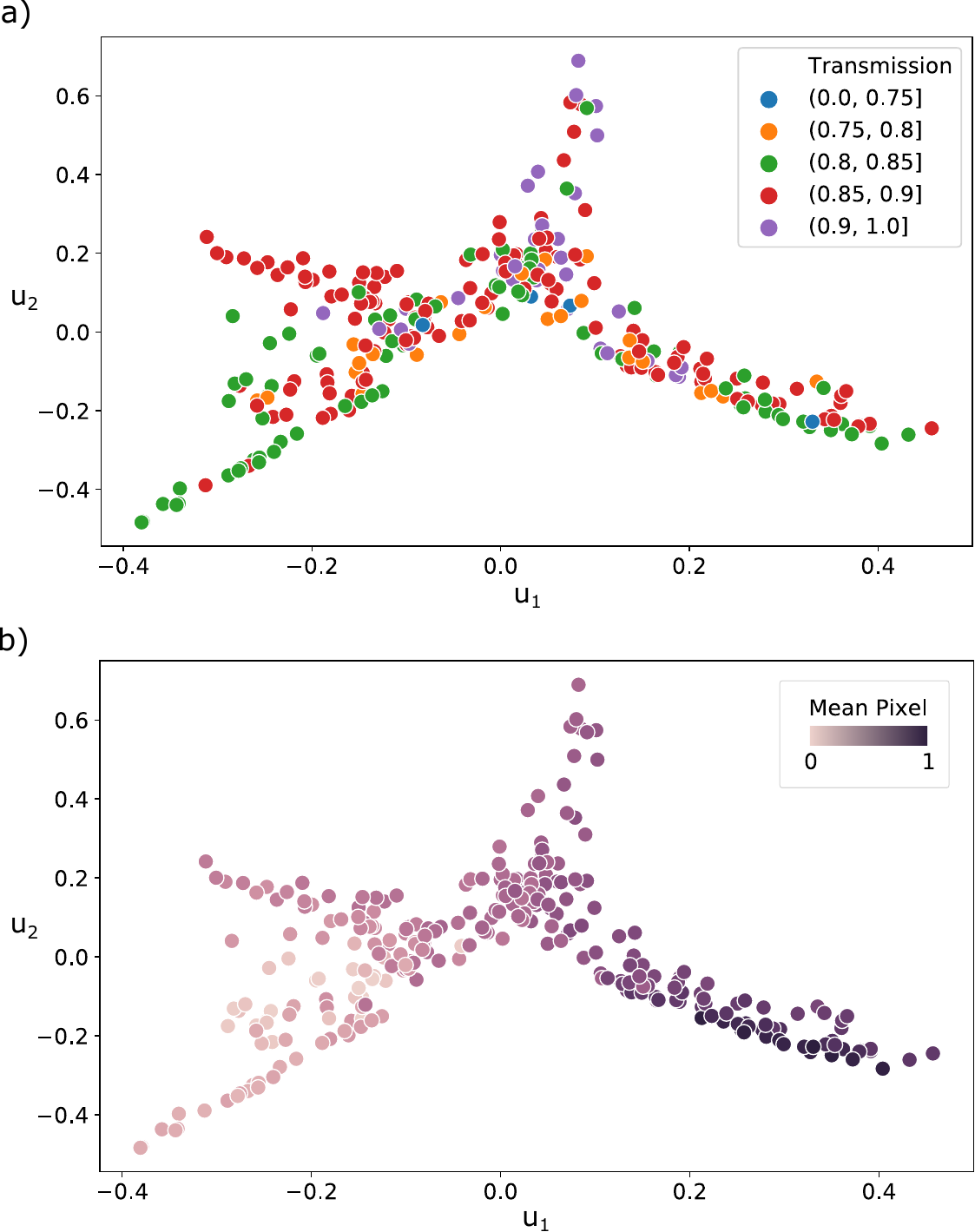}
    \caption{Spectral embedding of the optimized continuous structures from Figure \ref{fig:wdm-plots}.  a) Spectral embedding colored binned by the transmission values at 1550 nm. $u_1$ and $u_2$ correspond to the values of the first and second embedding vector. b) Same spectral embedding as in (a) but colored by the mean pixel value of the initial condition. }
    \label{fig:landscape-cont}
\end{figure*}

In the main text, we presented the local minima encountered after running the discrete optimization. By repeating the same analysis but for the continuous permittivity structures, we see that the "discrete local minima" is actually determined by the continuous minima (Figure \ref{fig:landscape-cont}). The same exact conclusions drawn from the discrete structure analysis can be drawn for the continuous structure analysis. This is not too surprising as the discrete structure typically closely resembles the continuous structure.

\end{document}